\documentclass[aps,pra,showpacs,twocolumn,nofootinbib]{revtex4-2}
\usepackage{bm}
\usepackage{amsmath}
\usepackage{amssymb}
\usepackage{slashed}
\usepackage{float}
\allowdisplaybreaks
\usepackage{subcaption}
\usepackage{dcolumn}
\usepackage{graphicx}

\usepackage{physics}
\usepackage{nicefrac}
\newcolumntype{w}[1]{D{.}{.}{#1}}
\newcommand{\Za}{Z\alpha}
\newcommand{\vare}{\varepsilon}
\newcommand{\balpha}{\vec{\alpha}}
\newcommand{\bnabla}{\vec{\nabla}}
\newcommand{\bgamma}{\vec{\gamma}}
\newcommand{\bsigma}{\vec{\sigma}}

\newcommand{\rqq}{{\rm q}}

\newcommand{\rp}{{\rm p}}

\newcommand{\rk}{{\rm k}}
\newcommand{\bfp}{{\vec p}}

\newcommand{\bfq}{{\vec q}}
\newcommand{\bfx}{{\vec x}}
\newcommand{\bfk}{{\vec k}}
\newcommand{\bfz}{{\vec z}}

\newcommand{\Dmatrix}[4]{
        \left(
        \begin{array}{cc}
        #1  & #2   \\
        #3  & #4   \\
        \end{array}
        \right)
        }

\begin{document}

\title{One-loop electron self-energy with
accelerated partial-wave expansion in Coulomb gauge}

\author{V.~A. Yerokhin}
\email{Corresponding author, vladimir.yerokhin@mpi-hd.mpg.de}
\affiliation{Max~Planck~Institute for Nuclear Physics, Saupfercheckweg~1, D~69117 Heidelberg, Germany}

\author{Z. Harman}
\affiliation{Max~Planck~Institute for Nuclear Physics, Saupfercheckweg~1, D~69117 Heidelberg, Germany}

\author{C.~H. Keitel}
\affiliation{Max~Planck~Institute for Nuclear Physics, Saupfercheckweg~1, D~69117 Heidelberg, Germany}

\begin{abstract}

Numerical calculations of the electron self-energy without any expansion in the binding nuclear field
are required in order to match the rapidly advancing precision of experimental spectroscopy.
For the lightest elements, particularly hydrogen, these computations are complicated
by large numerical cancelations and the slow convergence of the partial-wave expansion.
Methods with accelerated convergence of the partial-wave expansion have been recently
put forward [V. A. Yerokhin, K. Pachucki, V. M. Shabaev, Phys.
Rev. A 72, 042502 (2005); J. Sapirstein and K. T. Cheng, Phys. Rev. A 108, 042804
(2023)]. In our work we extend the accelerated-convergence methods to the previously
hardly accessible region of nuclear charges $Z < 5$ and higher excited states.

\end{abstract}

\maketitle

The electron self-energy is the leading quantum electrodynamics (QED) contribution to atomic energy levels. To match the precision of modern experiments \cite{loetzsch:24,pfafflein:24,gassner:18,kraft:17}, this effect must be calculated
 \cite{indelicato:19}
 with high accuracy and without expansion in the nuclear binding strength parameter $\Za$, where
$Z$ is the nuclear charge number and $\alpha$ is the fine-structure constant.

Numerical calculations of the self-energy correction
to all orders in $\Za$ have a long history.
The first such calculation was carried out by Desiderio and Johnson \cite{desiderio:71}, based on the method proposed by Brown, Langer, and Schaefer \cite{brown:59:procI}.
A real break-through, however, was achieved later by Peter Mohr \cite{mohr:74:a,mohr:74:b,mohr:82,mohr:92:b},
who performed accurate calculations across a broad range of ions and electron states. Despite this progress, calculations for a small range of
$Z$ values, specifically $Z<5$,
remained inaccessible for some time.
It took nearly two decades to bridge this gap and extend calculations to the lightest elements,
including hydrogen \cite{jentschura:99:prl,jentschura:01:pra,jentschura:04:se}.
These calculations required heroic efforts, the use of multiprecision arithmetic, and the inclusion of
partial-wave expansion contributions as high as several millions.

Several other methods have been reported in
the literature for calculations of the electron self-energy
\cite{indelicato:92:se,persson:93:ps,quiney:93,labzowsky:97:mcm,indelicato:98,zamastil:12}.
The most widely used method, however, was introduced  by Snyderman \cite{snyderman:91}
and implemented  by Blundell and Snyderman \cite{blundell:91:se}. This method,
sometimes referred to as the potential expansion approach,
was quickly adopted by other groups \cite{cheng:93,mitrushenkov:95,yerokhin:99:pra}
and, most importantly, was successfully generalized for evaluations of
higher-order self-energy corrections
\cite{persson:96:2el,blundell:97:pra,yerokhin:99:sescr,yerokhin:00:lalpra}.

The potential-expansion approach produces  results that are
typically less accurate
than those obtained by Mohr and collaborators. The main factor limiting the numerical accuracy
of this approach is
the uncertainty arising from the extrapolation of the partial-wave (PW) expansion
of the electron propagator inside the self-energy loop.
Recently it was realized that it is possible
to accelerate the convergence of the PW expansion and thereby
increase  the accuracy achievable in this method by orders of magnitude.
Specifically, there were approaches proposed by Yerokhin, Pachucki, and Shabaev (YPS)
\cite{yerokhin:05:se} and by Sapirstein and Cheng (SC)
\cite{sapirstein:23}. They will be referred to as the YPS and SC
accelerated-convergence schemes, respectively.

Both the YPS and SC approaches are based on
identities  that commute the Coulomb potentials outside the free-electron propagators in the
potential expansion of the Dirac-Coulomb Green function.
The YPS approach was extensively used in practical calculations
\cite{yerokhin:11:fns,yerokhin:11:prl,yerokhin:20:gfact}, but only for the
first-order self-energy matrix elements and its derivatives. Extending this method to
higher-order self-energy diagrams has proven to be difficult and never has been demonstrated so far.
In contrast, the SC convergence-acceleration
approach was already generalized to the vertex corrections
\cite{malyshev:24} and to the two-loop self-energy \cite{yerokhin:24:tobe}.

Both the YPS and SC approaches have mostly reported results for $Z \ge 5$,
due to large numerical cancellations that arise for lighter elements.
The cause of these cancellations is well understood. Snyderman \cite{snyderman:91}
demonstrated that individual self-energy contributions in Feynman gauge contain spurious
terms of order $\alpha(\Za)^2$, which cancel out in the sum to yield the physical result
of order $\alpha(\Za)^4$
and thus cause numerical cancelations for low $Z$.
As a potential solution, Snyderman
suggested using the Fried-Yennie (FY) gauge, in which these spurious terms are absent.
Another possibility to avoid numerical cancellations is to use the Coulomb gauge
\cite{adkins:83,adkins:86,hedendahl:12}.

In the present work we perform calculations of the electron self-energy
in different gauges and demonstrate that the Coulomb gauge is the optimal choice
for calculations in the low-$Z$ region. Furthermore, we implement
both the YPS and SC accelerated-convergence schemes in the Coulomb gauge and compare their
performance. As a result, we establish a calculation scheme that is applicable across
the entire $Z$ region including hydrogen, and arbitrary excited reference states.

The paper is organized as follows. In Sec.~\ref{sec:1} we describe the
generalization of the potential-expansion method for the cases of the
general covariant gauge and the Coulomb gauge. This description
is also the basis of the accelerated-convergence approaches.
In Sec.~\ref{sec:2} we discuss the YPS and SC accelerated-convergence
schemes in the Coulomb gauge.
In Sec.~\ref{sec:3} we describe details of our numerical implementation.
Sec.~\ref{sec:4} presents our
numerical results and discussion.

The relativistic units ($\hbar=c=m=1$) and the Heaviside charge units ($ \alpha = e^2/4\pi$, $e<0$)
are used throughout this paper.
We use roman style ($\rp$) for four vectors, an explicit vector notation ($\bfp$) for three
vectors and italic style ($p$) for scalars.
Four vectors have the form $\rp = (\rp_0,\bfp)$.

\section{Potential-expansion approach}
\label{sec:1}

\subsection{Basic formulas}
The unrenormalized expression for the one-loop electron self-energy correction
to an energy level of a bound state $a$ is
\begin{align} \label{lamb1}
\Delta E_{\rm SE,nren} & \ = 2i\alpha \int_{C_F} d\omega
  \int d^3x_1
         d^3 x_2\,
         \psi^{\dag}_a(\bfx_1)\,
          \alpha_{\mu}\,
        \nonumber \\ &
  \times
        G(\vare_a-\omega,\bfx_1,\bfx_2)\,
         \alpha_{\nu}\,
        \psi_a(\bfx_2)\, D^{\mu \nu}(\omega, \bfx_{1},\bfx_2)\,,
\end{align}
where $\alpha_{\mu} = (1,\balpha)$, $\balpha$ are the Dirac matrices,
$G(\omega,\bfx_1,\bfx_2)$ is the Dirac-Coulomb
Green function,
$D^{\mu \nu}(\omega, \bfx_{1},\bfx_2)$ denotes the photon propagator (see Appendix~\ref{app:phot}),
and $C_F$ is the standard Feynman integration contour.
$\psi_a(\bfx)$ is the reference-state wave function,
which is a bound solution of the Dirac equation with the
energy $\vare_a$ and has the form
\begin{equation} \label{wfx}
\psi_{a}(\bfx)
        =\left(  {g_a(x)\, \chi_{ \kappa_a  \mu_a}(\hat{\bfx})}
           \atop{if_a(x)\, \chi_{-\kappa_a \mu_a}}(\hat{\bfx}) \right) \,,
\end{equation}
where $g_a$ and $f_a$ are the radial components,
$\chi_{\kappa\mu}$ are the Dirac spin-angular spinors \cite{rose:61}, $x = |\bfx|$,
and $\hat{\bfx} = \bfx/x$.

It is well known that the unrenormalized expression given by Eq.~(\ref{lamb1}) is ultraviolet
(UV) divergent and requires renormalization. Specifically, one has to regularize
the UV divergencies (preferably, in a covariant manner), subtract
the mass counter-term contribution, eliminate all divergent terms analytically,
and obtain explicitly finite expressions that can be evaluated numerically.

The
potential-expansion approach for the evaluation of the one-loop self-energy correction was
introduced by Snyderman in Ref.~\cite{snyderman:91}; the detailed description
of this method was also presented in Ref.~\cite{yerokhin:99:pra}. In this work, we concentrate on
the generalization of the scheme of Ref.~\cite{yerokhin:99:pra}
for the cases of the general covariant gauge and the Coulomb gauge.

In the potential-expansion method,
the self-energy correction after renormalization is represented as a sum of the
zero-potential, one-potential, and many-potential contributions,
\begin{equation} 
\Delta E_{\rm SE} = \Delta E^{(0)}_{\rm SE}+ \Delta E^{(1)}_{\rm SE} + \Delta E^{(2+)}_{\rm SE}\,,
\end{equation}
where the upper index indicates the number of interactions with the binding
Coulomb field inside the self-energy loop. In the following, we will consider each of
these contributions in turn.

\subsection{Zero-potential term}

The zero-potential term is represented as (see Eq.~(11) of Ref.~\cite{yerokhin:99:pra})
\begin{equation}\label{eq:1}
\Delta E^{(0)}_{\rm SE} = \int \frac{d^3p}{(2\pi)^3}\,
        \overline{\psi}_a(\bfp)\, \Sigma^{(0)}_R(\vare_a,\bfp)\, \psi_a(\bfp) \, ,
\end{equation}
where $\overline{\psi} = \psi^{\dag}\gamma^0$,
$\psi_a(\bfp)$ is the reference-state
wave function in momentum space, and $\Sigma^{(0)}_R(\vare,\bfp)$ is the renormalized
one-loop free-electron self-energy operator. The wave function in momentum space is defined as
\begin{equation} \label{wfp}
\psi_a(\bfp) = \int d^3 x\,
        e^{-i\bfp\cdot\bfx}\, \psi_a(\bfx)
=
i^{-l_a}
\left( {g_a(p)\, \chi_{\kappa_a \mu_a}(\hat{\bfp})}
        \atop{f_a(p)\, \chi_{-\kappa_a \mu_a}}(\hat{\bfp}) \right) \,,
\end{equation}
where
$l_a = |\kappa_a+1/2|-1/2$.

In the general covariant gauge,
the renormalized one-loop free-electron self-energy operator has the form similar to that in the Feynman gauge
(cf. Eq.~(A5) of Ref.~\cite{yerokhin:99:pra}),
\begin{equation}
\Sigma^{(0)}_R(\vare,\bfp) = \frac{\alpha}{4\pi}\,\Big[a(\rho) + \slashed{p}\,b(\rho)\Big]\,,
\end{equation}
where $\slashed{p} = p_{\mu}\gamma^{\mu} = \vare\gamma^0 - \bgamma\cdot\bfp$,
$\rho = 1-\vare^2+\bfp\,^2$, and the functions $a(\rho)$ and $b(\rho)$ are defined in Appendix~\ref{sec:freese1}.

In the Coulomb gauge, the free-electron self-energy operator becomes more complex. The corresponding expression
was derived by Adkins \cite{adkins:83} and has the form
\begin{align}
\Sigma^{(0)}_{R,\rm Coul}(\vare,\bfp) =&\  \frac{\alpha}{4\pi}
 \int_0^1 dx\, du\,
  \Big[
  	a(p)
   + \gamma^0\vare\,b(p) + \bgamma\cdot\bfp\, c(p)
 \Big]\,,
\end{align}
where $x$ and $u$ are Feynman parameters and formulas for the functions $a$, $b$, and $c$ are given in
Appendix~\ref{sec:freese2}.

The angular integration over $\hat{\bfp}$ in Eq.~(\ref{eq:1}) is easily performed
analytically by noting that
\begin{align}\label{eq:3}
\slashed{p} = \vare\gamma_0 - \bgamma\cdot\bfp = \Dmatrix{\vare}{-\bsigma\cdot\bfp}{\bsigma\cdot\bfp}{-\vare}\,,
\end{align}
using the identity
\begin{align}\label{eq:4}
\bsigma\cdot\hat{\bfp}\,\chi_{\kappa\mu}(\hat{\bfp}) = -\chi_{-\kappa\mu}(\hat{\bfp})\,,
\end{align}
and the normalization condition
\begin{align}
\int d\hat{\bfp}\, \chi_{\kappa\mu}^{\dag}(\hat{\bfp})\,\chi_{\kappa\mu}(\hat{\bfp}) = 1\,.
\end{align}
The resulting expression for the zero-potential term
in Coulomb gauge is
\begin{align}
\Delta E^{(0)}_{\rm SE, Coul} = &\ \frac{\alpha}{32\pi^4}\,
 \int_0^{\infty}dp\,
   p^2\,
 \int_0^1 dx\, du\,
  \Big[
   {a(p)}\,(g_a^2-f_a^2)
\nonumber \\   &
    + {b(p)}\,\vare_a(g_a^2+f_a^2) -c(p)\,2 p\,  g_af_a
 \Big]\,,
\end{align}
where $g_a \equiv g_a(p)$ and $f_a \equiv f_a(p)$.
We note that the integration over one Feynman parameter ($u$) can easily be
performed analytically. However, we prefer to evaluate the above expression
as it stands, since it is computationally very cheap and can be performed to an arbitrary accuracy.

\subsection{One-potential term}

The one-potential term  reads (see Eq.~(12) of Ref.~\cite{yerokhin:99:pra})
\begin{equation} \label{eq:2}
\Delta E_{\rm SE}^{(1)} =
        \int \frac{d^3\bfp\,^{\prime}}{(2\pi)^3} \frac{d^3\bfp}{(2\pi)^3}\,
        \overline{\psi}_a(\bfp\,^{\prime})\, \Gamma_R^{\,0}(\rp^{\prime},\rp)\,
        V(q)\, \psi_a(\bfp) \, ,
\end{equation}
where $q = |\bfq|$, $\vec{q} = \bfp\,^{\prime}-\bfp$, $V(q) = -4\pi\Za/q^2$ is the
binding nuclear potential in momentum space, $\Gamma_R^{\,0}$ is
the time component of the renormalized free-electron vertex operator
$\Gamma^{\mu}_R$, and the time components of the 4-vectors $\rp^{\prime}$ and $\rp$
are fixed by $p^{\prime}_0 = p_0 = \vare_a$.

In the general covariant gauge, the free-electron vertex operator has a form
\begin{align}
\Gamma_R^{\,0}(\rp^{\prime},\rp) =&\ \frac{\alpha}{4\pi}\int_0^1dy\,
 \Big[
 \gamma^0A + \slashed{p}^{\prime}\,B + \slashed{p}\,\,C
  + \slashed{p}^{\prime}\gamma^0\slashed{p}\,D
\nonumber \\ &
 + \slashed{p}^{\prime}\gamma^0\,E  + \gamma^0\slashed{p}\,F + \slashed{p}^{\prime}\slashed{p}\, G+ H
 \Big]\,,
\end{align}
where
the scalar functions $A$-$H$ are obtained from formulas in Appendix~\ref{sec:freever1}.
The above expression is analogous to the Feynman-gauge representation (see Eq.~(B3) of
Ref.~\cite{yerokhin:99:pra}), with additional functions $E$, $F$, and $G$ that are absent in the Feynman gauge.

In the Coulomb gauge, the expression for the free-electron vertex operator becomes more complicated. The derivation
of Adkins \cite{adkins:86} yields
\begin{align}
\Gamma^{\,0}_{R,\rm Coul}(\rp^{\prime},\rp) =&\
 \frac{\alpha}{4\pi}
  \int_0^1 dx\,du\,
   \Big[
   \gamma^0{\cal A}
   + \bgamma\cdot\bfp\,^{\prime}\,{\cal B}
\nonumber \\&
   + \bgamma\cdot\bfp\,\,{\cal C}
   + \bgamma\cdot\bfp\,^{\prime} \gamma^0 \bgamma\cdot\bfp\,\,{\cal D}
\nonumber \\&
   + \bgamma\cdot\bfp\,^{\prime}\gamma^0\,{\cal E}
   + \gamma^0\bgamma\cdot\bfp\,\,{\cal F}
   + {\cal H}
   \Big]\,,
\end{align}
where $x$ and $u$ are Feynman parameters and scalar functions ${\cal A}$-${\cal H}$ are
defined in Appendix~\ref{sec:freever2}. Note that there are only two Feynman parameters here;
the third Feynman parameter $s$ is assumed to be integrated out analytically.

In Eq.~(\ref{eq:2}) we need to analytically perform integrations over all angular variables except
$\xi = \hat{\bfp}\,^{\prime}\cdot \hat{\bfp}$. The simplest way to do this is to average over the
momentum projections of the reference state $\mu_a$, using the fact that energy corrections
do not depend on $\mu_a$. After that, the action of $\slashed{p}$ and $\bgamma\cdot\bfp$ on the
reference-state wave functions is worked out with help of Eqs.~(\ref{eq:3}) and (\ref{eq:4})
and the spin-angular spinors are simplified by the identity
\begin{align}
\frac{4\pi}{2j+1}\sum_{\mu}
 \chi_{\kappa\mu}^{\dag}(\hat{\bfp}\,^{\prime})\,\chi_{\kappa\mu}(\hat{\bfp}) =
P_{|\kappa+1/2|-1/2}(\xi)\,,
\end{align}
where $P_l$ is the Legendre polynomial.

It is convenient to introduce the integrals $X[Y]$ of basis angular structures $Y$ as follows
\begin{align}
\frac{1}{2j_a+1}&\ \sum_{\mu_a}
\int d\hat{\bfp}\,^{\prime} d\hat{\bfp}\,\, \overline{\psi}_a(\bfp\,^{\prime})\, Y\, F(\xi)\,\psi_a(\bfp)
\nonumber \\ &
= \frac1{4\pi}\,\int d\hat{\bfp}\,^{\prime}  d\hat{\bfp}\,\, X[Y] \, F(\xi)
=  2\pi\int_{-1}^1d\xi\, X[Y] \, F(\xi)\,,
\end{align}
where $F(\xi)$ is an arbitrary function of $\xi$. We obtain
\begin{align}
X[\gamma^0] = &\ g_a^{\prime}g_a\,P_{l_a} + f_a^{\prime}f_a\,P_{\overline{l}_a}\,,\nonumber \\
X[1]        = &\ g_a^{\prime}g_a\,P_{l_a} - f_a^{\prime}f_a\,P_{\overline{l}_a}\,,\nonumber \\
X[\bgamma\cdot\bfp\,]        = &\ -p\, g_a^{\prime}f_a\,P_{l_a} - p\, f_a^{\prime}g_a\,P_{\overline{l}_a}\,,\nonumber \\
X[\gamma^0\bgamma\cdot\bfp\,]= &\ -p\, g_a^{\prime}f_a\,P_{l_a} + p\, f_a^{\prime}g_a\,P_{\overline{l}_a}\,,\nonumber \\
X[\bgamma\cdot\bfp\,^{\prime}\,]     = &\ -p^{\prime}\, f_a^{\prime}g_a\,P_{l_a} - p^{\prime}\, g_a^{\prime}f_a\,P_{\overline{l}_a}\,,\nonumber \\
X[\bgamma\cdot\bfp\,^{\prime}\gamma^0]= &\ -p^{\prime}\, f_a^{\prime}g_a\,P_{l_a} + p^{\prime}\, g_a^{\prime}f_a\,P_{\overline{l}_a}\,,\nonumber \\
X[\bgamma\cdot\bfp\,^{\prime}\gamma^0\bgamma\cdot\bfp\,]= &\ p^{\prime}p\, f_a^{\prime}f_a\,P_{l_a} + p^{\prime}p\, g_a^{\prime}g_a\,P_{\overline{l}_a}\,,
\label{eq:5}
\end{align}
where $g_a^{\prime}\equiv  g_a(p^{\prime})$, $f_a^{\prime}\equiv  f_a(p^{\prime})$,
$g_a\equiv  g_a(p)$, $f_a\equiv  f_a(p)$, $P_l\equiv P_l(\xi)$, $l_a = |\kappa_a+1/2|-1/2$,
and $\overline{l}_a = |\kappa_a-1/2|-1/2$.

With help of Eq.~(\ref{eq:5}), the angular integration in the one-potential term is easily carried out.
The resulting expression for the one-potential term in the Coulomb gauge reads
\begin{align}\label{eq:5b}
\Delta E_{\rm SE, Coul}^{(1)} &\ =
-\frac{Z\alpha^2}{32\pi^5} \int_0^{\infty}dp^{\prime}dp\,
 \int_{-1}^{1}d\xi\,
   \int_0^1 dx\,du\,
  \frac{p^{\prime 2}p^2}{q^2}\,
\nonumber \\&
 \times
   \Big\{
   X[\gamma^0]{\cal A}
   + X[\bgamma\cdot\bfp\,^{\prime}\,]{\cal B}
   + X[\bgamma\cdot\bfp\,]\,{\cal C}
\nonumber \\&
   + X[\bgamma\cdot\bfp\,^{\prime} \gamma^0 \bgamma\cdot\bfp\,]\,{\cal D}
   + X[\bgamma\cdot\bfp\,^{\prime}\gamma^0]\,{\cal E}
\nonumber \\&
   + X[\gamma^0\bgamma\cdot\bfp]\,\,{\cal F}
   + X[1]\,{\cal H}
 \Big\}\,.
\end{align}
The one-potential term in the general covariant gauge is worked out in the complete analogy
with that for the Coulomb gauge.

The above expression for the one-potential term can already be calculated numerically. However,
its computation is complicated by the presence of the integrable singularity of the Coulomb
potential at $q\to 0$. In order to simplify numerical integrations, we subtract from the
vertex operator $\Gamma^{\,0}_R(\rp^{\prime},\rp)$ its diagonal in $p$ contribution, which
weakens the Coulomb singularity. The subtracted contribution can then be readily evaluated
by performing one momentum integration analytically. More specifically,
we transform Eq.~(\ref{eq:2}) as
\begin{align} \label{eq:6}
\Delta E_{\rm SE}^{(1)} =&
        \int \frac{d^3\bfp\,^{\prime}d^3\bfp}{(2\pi)^6}
        \overline{\psi}_a(\bfp\,^{\prime})
         \Big[\Gamma_R^{\,0}(\rp^{\prime},\rp)
        - \Gamma_R^{\,0}(\rp,\rp)\Big]V(q)
         \psi_a(\bfp)
\nonumber \\ &
        + \int \frac{d^3\bfp}{(2\pi)^3} \,
        \overline{\psi}_{Va}(\bfp)\,
        \Gamma_R^{\,0}(\rp,\rp)\,
        \psi_{a}(\bfp)
        \,,
\end{align}
where $\psi_{Va}(\bfp)$ is the Fourier transform of the product of the Coulomb potential and the wave function,
\begin{equation} \label{wfVa}
\psi_{Va}(\bfp) = \int d^3 x\,
        e^{-i\bfp\cdot\bfx}\, V(x)\,\psi_a(\bfx)\,.
\end{equation}
A similar transformation weakening the Coulomb singularity in the one-potential term
was used also in previous investigations, notably Refs.~\cite{mitrushenkov:95,sapirstein:23}.
We note that the Fourier transform of the product of the Coulomb potential and the wave function,
$\psi_{Va}(\bfp)$, can be readily obtained analytically, see Ref.~\cite{mohr:98} for
explicit formulas.

Even after the subtraction of the diagonal in $p$ contribution, the first term in Eq.~(\ref{eq:6})
still contain some singularity at small $q$ which complicates its numerical evaluation. In order
to handle it, it convenient to make the following change of variables \cite{yerokhin:99:pra}
$(p^{\prime},p,\xi) \to (x,y,q)$, where $x = p+p^{\prime}$, $y = |p-p^{\prime}|$, and
$q^2 = p^2+p^{\prime2}-2pp^{\prime}\xi$. This transforms the integral over $(p^{\prime},p,\xi)$ as follows
\begin{align} \label{eq:7}
\int_0^{\infty}dp^{\prime}dp\, &\ \int_{-1}^{1}d\xi\,F(p^{\prime},p,\xi) =
\int_0^{\infty}dx
 \int_0^x dy
\nonumber \\ & \times
 \int_y^x dq\,
 \frac{q}{2p^{\prime}p}
 \Big[F(p^{\prime},p,\xi) + F(p,p^{\prime},\xi)\Big]\,.
\end{align}
Note the appearance of the $q$ factor in the numerator as the consequence of the variable transformation,
which softens the behavior of the integrand at small $q$.

\subsection{Many-potential term}

The many-potential term $\Delta E^{(2+)}_{\rm SE}$ is obtained from Eq.~(\ref{lamb1}) by applying the
substitution $G(E,\bfx_1,\bfx_2) \to G^{(2+)}(E,\bfx_1,\bfx_2)$, where $G^{(2+)}(E,\bfx_1,\bfx_2)$
is the Dirac-Coulomb Green function containing two or more interactions with the
binding field. In practical calculations $G^{(2+)}$ is often represented as
\begin{align}
G^{(2+)}(E,\bfx_1,\bfx_2) =&\ G(E,\bfx_1,\bfx_2) - G^{(0)}(E,\bfx_1,\bfx_2)
\nonumber \\ &
 - G^{(1)}(E,\bfx_1,\bfx_2)\,,
\end{align}
where $G^{(0)}(E,\bfx_1,\bfx_2)$ is the free-electron Dirac Green function and $G^{(1)}(E,\bfx_1,\bfx_2)$
is the one-potential Dirac Green function defined as
\begin{align}\label{eq:G1}
G^{(1)}(E,\bfx_1,\bfx_2) = \int d^3z\,G^{(0)}(E,\bfx_1,\bfz)\, V(z)\,G^{(0)}(E,\bfz,\bfx_2)\,,
\end{align}
where $V(x) = -\Za/x$ is the Coulomb potential.

We now consider the many-potential term in the general covariant gauge. The general-gauge photon propagator $D^{\mu\nu}$
can be represented (see Appendix~\ref{app:phot}) as the sum of the Feynman-gauge propagator $D_F^{\mu\nu}$
and the gauge-dependent term $D_{\xi}^{\mu\nu}$,
\begin{align}
D^{\mu\nu}(\omega,\bfx_1,\bfx_2) = D_F^{\mu\nu}(\omega,\bfx_1,\bfx_2) + \xi\,D_{\xi}^{\mu\nu}(\omega,\bfx_1,\bfx_2)\,,
\end{align}
where $\xi$ is the gauge parameter.
Consequently, the many-potential term is divided into two parts,
\begin{align}
\Delta E^{(2+)}_{\rm SE} = \Delta E^{(2+)}_{F} + \xi \Delta E^{(2+)}_{\xi}\,.
\end{align}
The Feynman-gauge many-potential part is described in detail in
Ref.~\cite{yerokhin:99:pra},
so here we concentrate on the gauge-dependent part.

\begin{widetext}

Using Eq.~(\ref{lamb1}), the substitution $G\to G^{(2+)}$, and formulas for the photon propagator from Appendix~\ref{app:phot},
we write the gauge-dependent many-potential part as
\begin{align}\label{many0}
\Delta E^{(2+)}_{\xi} = &\ 2i\alpha \int_{C_F}d\omega\,
  \int d^3x_1\,d^3x_2\,
   \int \frac{d^3k}{(2\pi)^3} \frac{-1}{(\omega^2-\bfk^2+i0)^2}\,
   \psi_a^{\dag}(\bfx_1)\,
 \nonumber \\ & \times
    \Big[ (\omega + i\balpha_1\cdot \bnabla_1) \, e^{i\bfk\cdot\bfx_1}\Big] \,
     G^{(2+)}(\vare_a-\omega,\bfx_1,\bfx_2)\,
    \Big[ (\omega - i\balpha_2\cdot \bnabla_2) \, e^{-i\bfk\cdot\bfx_2}\Big] \,
    \psi_a(\bfx_2)\,,
\end{align}
where gradients are supposed to act only on the exponential functions in the brackets. We now replace
$\balpha \cdot \bnabla$ acting on a function
by the commutator of the function with
the Dirac Hamiltonian $h_D$, $h_D(x) = -i\balpha\cdot\bnabla + \beta\,m + V(x)$,
\begin{align}
-i \balpha \cdot \bnabla \, e^{i\bfk\cdot\bfx} = \Big[ h_D(x),e^{i\bfk\cdot\bfx}\Big]\,
\end{align}
and evaluate the commutator,
letting the Hamiltonian
act on the wave function $\psi_a$ (producing the reference-state energy $\vare_a$) and
the Green function.
We obtain
\begin{align}\label{many1}
\Delta E^{(2+)}_{\xi} = &\ 2i\alpha \int_{C_F}d\omega\,
  \int d^3x_1\,d^3x_2\,
   \psi_a^{\dag}(\bfx_1)\,
 \nonumber \\ & \times
    \Big[ (\omega -\vare_a + h_D(x_1))  \,
     G^{(2+)}(\vare_a-\omega,\bfx_1,\bfx_2)\,
    (\omega -\vare_a + h_D^{\dag}(x_2)) \Big] \,
    \psi_a(\bfx_2)
    \dot{D}(\omega,x_{12})\,
    \,,
\end{align}
where $\dot{D}(\omega) = -1/(2\omega) \partial/(\partial \omega) D(\omega)$,
$D(\omega)$ is defined by Eq.~(\ref{ph0}), and both Hamiltonians act on the Green function.
The action of $h_D$ on the Green functions is as follows
\begin{align}
(\vare-h_D(x_1))\,G(\vare,\bfx_1,\bfx_2) = &\ \delta^3(\bfx_1-\bfx_2)\,, \nonumber \\
(\vare-h_D(x_1))\,G^{(0)}(\vare,\bfx_1,\bfx_2) = &\ \delta^3(\bfx_1-\bfx_2)
  - V(x_1)\,G^{(0)}(\vare,\bfx_1,\bfx_2)\,\,, \nonumber \\
(\vare-h_D(x_1))\,G^{(1)}(\vare,\bfx_1,\bfx_2) = &\
   V(x_1)\,G^{(0)}(\vare,\bfx_1,\bfx_2)
   - V(x_1)\,G^{(1)}(\vare,\bfx_1,\bfx_2)
   \,\,. \nonumber \\
\end{align}
Therefore,
\begin{align}
(\vare-h_D(x_1))\,G^{(2+)}(\vare,\bfx_1,\bfx_2) = &\
    V(x_1)\,G^{(1)}(\vare,\bfx_1,\bfx_2)
   \,,
\end{align}
and, consequently,
\begin{align}
(\vare-h_D(x_1))\,G^{(2+)}(\vare,\bfx_1,\bfx_2)\,(\vare-h_D^{\dag}(x_2)) = &\
    V(x_1)\,\Big[G^{(0)}(\vare,\bfx_1,\bfx_2)
    - G^{(1)}(\vare,\bfx_1,\bfx_2)\Big] \, V(x_2)
   \,.
\end{align}
We thus obtain a very simple representation for the gauge-dependent part of the many-potential term in the general covariant gauge,
\begin{align}\label{many2}
\Delta E^{(2+)}_{\xi} = &\ 2i\alpha \int_{C_F}d\omega\,
  \int d^3x_1\,d^3x_2\,
   \psi_a^{\dag}(\bfx_1)\,
    V(x_1)\,\Big[G^{(0)}(\vare_a-\omega,\bfx_1,\bfx_2)
    - G^{(1)}(\vare_a-\omega,\bfx_1,\bfx_2)\Big] \, V(x_2)\,
    \psi_a(\bfx_2)\,
    \dot{D}(\omega,x_{12})
    \,.
\end{align}

Now we turn to the evaluation of the many-potential term in the Coulomb gauge.
An expression similar to Eq.~(\ref{many2}) can be derived also for the Coulomb gauge.
However, it has the disadvantage that it cannot be easily generalized to the
case of the accelerated convergence schemes, where the many-potential Green function
$G^{(2+)}$ is substituted by a more complex function.
For this reason, we now perform the angular reduction of the many-potential term
in the Coulomb gauge in a
straightforward manner.

The simplest way to do this is to start with the unrenormalized expression (\ref{lamb1})
and employ the representation of the Green function as a sum over the Dirac spectrum (see, e.g.,
Ref.~\cite{yerokhin:20:green}). Then the summation over the momentum projections
is easily carried out (see Sec.~4 of Ref.~\cite{yerokhin:20:green}) and we obtain
\begin{align}
\Delta E_{\rm SE,nren} = \frac{i\alpha}{2\pi}
 \int_{C_F}d\omega\, \sum_{n,J}
  \frac{(-1)^{j_a+j_n+J}}{2j_a+1}\,
    \frac{R_J(\omega,anna)}{\vare_a-\omega-\vare_n}\,,
\end{align}
where $R_J$ are the standard two-body radial integrals, the same as for the one-photon exchange
correction, which are very well studied in the literature. In particular, in the
Coulomb gauge, expressions for the radial integrals $R_J$ were derived in
Ref.~\cite{mann:71}. We here adopt the expressions
for the Coulomb-gauge
radial integrals from Ref.~\cite{yerokhin:21:gfact} (see Appendix B in there).
Now we just rewrite those expressions in terms of the Green function
and then apply the substitution $G \to G^{(2+)}$, converting the unrenormalized
expression into the many-potential term.
We thus obtain
the many-potential term in the Coulomb gauge as
\begin{align} \label{many5}
\Delta E^{(2+)}_{\rm SE,Coul} =&\ \frac1{2j_a+1}
\frac{i\alpha}{2\pi}
         \int_{C_F} d\omega \,
\sum_{\kappa_n,J} 2
         \int_0^{\infty} dx_1 \int_0^{x_1} dx_2\,
  (x_1 x_2)^2
         \nonumber \\ & \times
        \bigg[ (2J+1) \left[ C_J(\kappa_n, \kappa_a) \right]^2
          g_J(0) \left\{ G^{(2+)}_{\kappa_n}(\vare_a-\omega) \right\}^I
  -\sum_{L=J-1}^{J+1}
  a_{JL}\,
  g_L(\omega) \left\{ G^{(2+)}_{\kappa_n}(\vare_a-\omega)
                           \right\}^{II}_{JLL}
         \nonumber \\ &
  + \sqrt{J(J+1)}\, g_J^{\rm ret,1}(\omega)
  \left\{ G^{(2+)}_{\kappa_n}(\vare_a-\omega)\right\}^{II}_{J,J+1,J-1}
  + \sqrt{J(J+1)}\, g_J^{\rm ret,2}(\omega)
  \left\{ G^{(2+)}_{\kappa_n}(\vare_a-\omega)\right\}^{II}_{J,J-1,J+1}
                                \bigg] \,,
\end{align}
where
the parentheses $\{\ldots\}^{I,II}$ are defined as follows
\begin{align} 
\Big\{ G(\vare) \Big\}^I =&\
          g_a(x_1)\, G^{11}(\vare,x_1,x_2)\, g_a(x_2)
                 +g_a(x_1)\, G^{12}(\vare,x_1,x_2)\, f_a(x_2) \nonumber \\
&
                 +f_a(x_1)\, G^{21}(\vare,x_1,x_2)\, g_a(x_2)
                 +f_a(x_1)\, G^{22}(\vare,x_1,x_2)\, f_a(x_2)\,,
\end{align}
\begin{align} 
\Big\{ G(\vare) \Big\}_{JLL'}^{II} =&\
         f_a(x_1)\,G^{11}(\vare,x_1,x_2)\, f_a(x_2)\,S_{JL}(-\kappa_a,\kappa_n)\, S_{JL'}(-\kappa_a,\kappa_n)
\nonumber \\ &
               - f_a(x_1)\,G^{12}(\vare,x_1,x_2)\, g_a(x_2)\,S_{JL}(-\kappa_a,\kappa_n)\, S_{JL'}(\kappa_a,-\kappa_n)
\nonumber \\ &
               - g_a(x_1)\,G^{21}(\vare,x_1,x_2)\, f_a(x_2)\,S_{JL}( \kappa_a,-\kappa_n)\, S_{JL'}(-\kappa_a,\kappa_n)
\nonumber \\ &
               + g_a(x_1)\,G^{22}(\vare,x_1,x_2)\, g_a(x_2)\,S_{JL}( \kappa_a,-\kappa_n)\, S_{JL'}( \kappa_a,-\kappa_n)\,,
\end{align}

\end{widetext}

\noindent
and the coefficients $a_{JL}$ are given by
\begin{align}
a_{JL} = \left\{
        \begin{array}{cl}
        J+1 \,,  & {\mbox{\rm for} \ \ } L = J-1 \,,\\
        2J+1 \,, & {\mbox{\rm for} \ \ } L = J \,,\\
        J   \,,  & {\mbox{\rm for} \ \ } L = J+1\,.
        \end{array}
        \right.
\end{align}
In the above, $C_{J}$ and $S_{JL}$ are the standard angular coefficients given, e.g., by Eqs.~(C7)-(C10) of Ref.~\cite{yerokhin:99:pra},
and $g_a(r)$ and $f_a(r)$ are the upper and lower radial components of the reference-state wave function.
Furthermore,
\begin{align} \label{eiomega}
g_l(0,x_1,x_2) =&\ \frac1{2l+1}\,\frac{x_<^l}{x_>^{l+1}}\,, \\
g_l(\omega, x_1 x_2) =&\ i\omega j_l(\omega x_<)\, h^{(1)}_l(\omega x_>) \,, \\
g^{\rm ret,1}_l(\omega,x_1,x_2)=  &\
         i\omega\,j_{l+1}(\omega x_<)\,h^{(1)}_{l-1}(\omega x_>) \,, \\
g^{\rm ret,2}_l(\omega,x_1,x_2)=  &\
         i\omega\,j_{l-1}(\omega x_<)\,h^{(1)}_{l+1}(\omega x_>)
            - \frac{\displaystyle 2l+1}{\displaystyle \omega^2}\frac{\displaystyle x_<^{l-1}}{\displaystyle x_>^{l+2}} \,,
\end{align}
where $x_< = \min(x_1,x_2)$, $x_> = \max(x_1,x_2)$,
$j_l(z)$, $h^{(1)}_l(z)$ are the spherical Bessel functions. Note that the function $g^{\rm ret,2}_l$ is regular
at $\omega \to 0$ despite the apparent singularity in the second term $\sim 1/\omega^2$. The disappearance of the
divergent term becomes evident when one makes the small-argument expansion of the spherical Bessel functions.
This cancellation, however, leads to numerical instabilities. In order to avoid them,
we used the regularized functions $\overline{j}_l$ and $\overline{h}^{(1)}_l$ for small $\omega$,
as described in
Ref.~\cite{yerokhin:21:gfact}. Note also that in Eq.~(\ref{many5}) we used the symmetry of the
integrand to restrict the integration region to $x_2\leq x_1$ only.

\begin{figure}[t]
\centerline{\resizebox{0.5\textwidth}{!}{\includegraphics{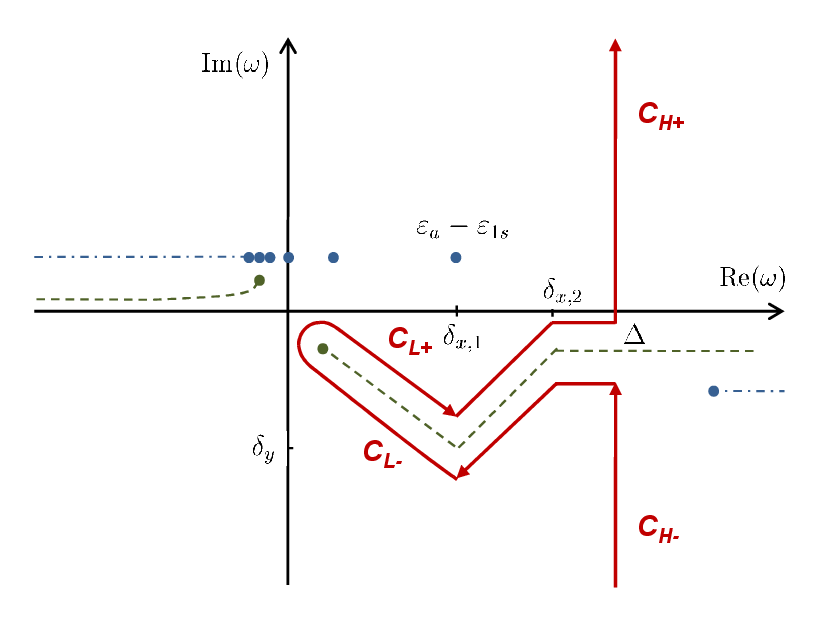}}}
 \caption{The poles and the branch cuts of the integrand of the matrix element of the self-energy
 operator and the integration contour $C_{LH}$ in the complex $\omega$ plane. The dashed lines
 (green) show the branch cuts of the photon
 propagator. The poles and the branch cuts of the electron propagator
 are shown by dots and the dashed-dot line (blue). The solid line (red) shows the integration
 contour $C_{LH}$.  \label{fig:CLH}}
\end{figure}

We now turn to discussing the optimal choice of the $\omega$ integration contour, which
is essential for the numerical evaluation of the many-potential term.
The standard Feynman integration contour in Eq.~(\ref{many5})
is not favorable for numerical calculations, since the Dirac Green function is a highly oscillating function for
large real $\omega$ and $x,x'\to \infty$. It is advantageous to deform the
integration contour into the region of large imaginary $\omega$ since the Dirac Green function
acquires an exponentially damping factor in there.

Fig.~\ref{fig:CLH} shows the contour $C_{LH}$ that we
used in our calculations, which is a modification
of the contour used by us earlier \cite{yerokhin:99:pra}.
It consists of the
low-energy part $C_L$ and the high-energy part $C_H$.
The low-energy part $C_L$ is bent into the complex plane
in order to avoid poles from the virtual intermediate states that are more deeply bound
than the reference state.
The low-energy part $C_L$ consists of two parts, $C_{L+}$ and
$C_{L-}$, the first of which runs on the upper bank of the cut of the photon propagator and the
second, on the lower bank and in the opposite direction.
Specifically, the contour $C_{L}$
consist of 3 sections: $(0,\delta_{x,1}-i\delta_y)$, $(\delta_{x,1}-i\delta_y,\delta_{x,2})$, and
$(\delta_{x,2},\Delta)$, each these sections are run twice in the opposite directions.
The high-energy part $C_H$ is parallel to the imaginary axis and consists of two parts, $C_{H-} =
(\Delta-i\infty, \Delta-i\epsilon)$ and $C_{H+} = (\Delta+i\epsilon, \Delta+i\infty)$. The
integrands for $C_{H+}$ and $C_{H-}$ are complex conjugated, so that it is sufficient to perform
the integration over $C_{H+}$ only and take twice the real part of the result.
A more detailed discussion of the integration contour $C_{LH}$ can be found in Sec.~5 of
Ref.~\cite{yerokhin:20:green}. We note also a very useful discussion of the analytical
properties of the self-energy in Ref.~\cite{mohr:74:a}.

For the numerical evaluation of radial integrals in Eq.~(\ref{many5}), we make the change of
variables proposed by Mohr \cite{mohr:82},
\begin{align}
\int_0^{\infty}dx_1\,&\ \int_0^{x_1}dx_2\,F(x_1,x_2)
\nonumber \\  &
 = \int_0^{\infty}dy\,
 \int_0^1dr\,\frac{y}{a^2}\,F\Big(\frac{y}{a},\frac{ry}{a}\Big)\,,
\end{align}
where $r = x_2/x_1$ and $y = a\,x_1$, with $a = 2 \sqrt{1-\vare_a^2}$.
We also use some of the integration prescriptions suggested in Ref.~\cite{mohr:74:b},
although with increased number of integration points.

\section{Accelerated-convergence approaches}
\label{sec:2}

We now examine methods to enhance the convergence of the partial-wave expansion
in the many-potential term.  The general idea is to find an approximation
for the many-potential
Green function $G^{(2+)}$ which captures
the slowest-converging part of the partial-wave expansion
{\em and} can be computed in a closed form.
Then we can subtract and re-add this approximation in the many-potential
term, separately calculating the subtracted contribution in a closed form,
without the partial-wave expansion.

There are two schemes that are able to achieve this goal,
both based on an idea initially proposed by Peter Mohr \cite{mohr:74:a}. Mohr
obtained an approximation for the one-potential Green function $G^{(1)}$
by using its integral representation and commuting the Coulomb
potential outside,
\begin{align}
G^{(1)}(\vare,\bfx_1,\bfx_2) &\ =
\int d^3z\, G^{(0)}(\vare,\bfx_1,\bfz)\, V(z)\,
G^{(0)}(\vare,\bfz,\bfx_2)
\nonumber \\
&\ \approx V(x_1)\,\int d^3z\, G^{(0)}(\vare,\bfx_1,\bfz)\,
G^{(0)}(\vare,\bfz,\bfx_2)
\nonumber \\ &
= - V(x_1)\,\frac{\partial}{\partial \vare}\, G^{(0)}(\vare,\bfx_1,\bfx_2)\,.
\end{align}
This simple approximation is based on the fact that the dominant
contribution to the self-energy integral comes from
the region where the radial arguments are close to each other,
$\bfx_1 \approx \bfx_2$.
In this region the commutators of the potential
$V$ with the free Green function $G^{(0)}$ are small and do
not contribute significantly to the partial-wave expansion.

Basing on the same idea, Sapirstein and Cheng \cite{sapirstein:23}
approximated the many-potential Green function $G^{(2+)}$ as
\begin{align}\label{acc2}
G^{(2+)}(\vare,\bfx_1,\bfx_2)
&\ \approx
 \frac12\, V(x_1)\,\frac{\partial^2}{\partial \vare^2}\, G^{(0)}(\vare,\bfx_1,\bfx_2)\,V(x_2)\,,
\end{align}
commuting the potentials outside in the two-potential term and
neglecting contributions with three or more potentials.

In earlier work by one of us \cite{yerokhin:05:se}, a similar approximation was
extended to all orders in the potential expansion, summing the entire series of multiple commutators.
The resulting approximation of that work is equivalent to
the following series
\begin{align}\label{acc3}
G^{(2+)}(\vare,\bfx_1,\bfx_2)
&\, \approx
 \frac12\, \Omega^2\,\frac{\partial^2}{\partial \vare^2}\, G^{(0)}(\vare,\bfx_1,\bfx_2)
\nonumber \\ &
+
 \frac1{6}\, \Omega^3\,\frac{\partial^3}{\partial \vare^3}\, G^{(0)}(\vare,\bfx_1,\bfx_2)
+ \ldots\,,
\end{align}
with $\Omega = 2\Za/(x_1+x_2)$.
It is clear that the first term in Eq.~(\ref{acc3}) agrees with
Eq.~(\ref{acc2}) in the limit $x_1\to x_2$  and that the next terms account for parts of the three- and higher-potential
contributions that are not included into Eq.~(\ref{acc2}).

We now consider the two accelerated-convergence approaches in detail.

\begin{widetext}

\subsection{YPS scheme}

We start with the approach introduced in Ref.~\cite{yerokhin:05:se} and referred to as the YPS scheme.
It uses the following approximation $G^{(2+)}_a$ of the many-potential Green function $G^{(2+)}$,
\begin{align}\label{13}
    G_{a}^{(2+)}(\vare,\bfx_1,\bfx_2)  =&\
G^{(0)}(\vare+\Omega,\bfx_1,\bfx_2)
   - G^{(0)}(\vare,\bfx_1,\bfx_2)
   -         \Omega\, \frac{\partial}{\partial \vare}\, G^{(0)}(\vare,\bfx_1,\bfx_2)\,,
\end{align}
with $\Omega = 2\Za/(x_1+x_2)$. Now we subtract and then re-add the function
$G_{a}^{(2+)}$ in the
many-potential self-energy term, representing it as a sum of two parts,
\begin{equation}\label{15}
    \Delta E_{\rm SE}^{(2+)} = \Delta E_{\rm acc, YPS}^{(2+)} + \Delta E_{\rm subtr, YPS}^{(2+)}\,.
\end{equation}
The first part $\Delta E_{\rm acc, YPS}^{(2+)}$ is obtained from $\Delta E_{\rm SE}^{(2+)}$
given by Eq.~(\ref{many5}) by applying the additional subtraction
$G^{(2+)} \to G^{(2+)}-G^{(2+)}_a$ to the parts that have slow convergence in partial waves.
Specifically, we apply this subtraction to the retarded-photon part of the high-energy contribution in
Eq.~(\ref{many5}). The low-energy contribution (corresponding to the $C_L$ part of the integration
contour) usually converges very fast and does not need any partial-wave extrapolation.
The instantaneous-photon part of the high-energy contribution (induced by the first term
in brackets in Eq.~(\ref{many5})) also converges fast and does not require acceleration.

The second term in Eq.~(\ref{15}), $\Delta E_{\rm subtr, YPS}^{(2+)}$, is the subtracted
contribution evaluated without partial-wave expansion. In the Feynman gauge, it was
worked out in Ref.~\cite{yerokhin:05:se}. Here we evaluate this contribution in the Coulomb
gauge. It reads (cf. Eq.~(19) of Ref.~\cite{yerokhin:05:se})
\begin{align}\label{16}
\Delta E_{\rm subtr, YPS}^{(2+)} &\ =
 -2i\alpha \int_{C_H}d\omega \,
  \int d^3x_1\,d^3x_2\,
    \psi_a^{\dag}(\bfx_1)\,
        \alpha_i\,
            G_{a}^{(2+)}(\vare_a-\omega,\bfx_1,\bfx_2)\,
            \alpha_j\,
                \psi_a(\bfx_2)\,
    D^{ij}_C(\omega,\bfx_1,\bfx_2)\,,
\end{align}
where $i,j = 1,2,3$. Using formulas (\ref{ph4}) and (\ref{ph5}) for the photon propagator,
the representation of $G_a^{(2+)}$ from Appendix~\ref{app:Ga},
and performing some algebra, we arrive at
\begin{align}
\Delta E_{\rm subtr, YPS}^{(2+)} &\ =
 2i\alpha \int_{C_H}d\omega \,
  \int d^3x_1\,d^3x_2\,
    \psi_a^{\dag}(\bfx_1)\,
\Big[
{\cal F}_1\,i\balpha\cdot\bfx_{12} + {\cal F}_2\,\beta + {\cal F}_3
\Big]
                \psi_a(\bfx_2)\,,
\end{align}
where the functions ${\cal F}_i = {\cal F}_i(x_1,x_2,x_{12})$ are given by
${\cal F}_1 = G_1\,(D_1+D_2)\,$,
${\cal F}_2 = G_2\,(3D_1-D_2)\,$, and
${\cal F}_3 = G_3\,(-3D_1+D_2)\,$, and the functions $D_{1,2}$ and $G_{1,2,3}$ are defined in
Appendices \ref{app:phot} and \ref{app:Ga}, respectively.
The integration over all angular variables except $\xi = \hat{\bfx}_1\cdot\hat{\bfx}_2$ is
carried out analytically in the same way as for the one-potential term.
We arrive at (cf. Eq.~(23) of Ref.~\cite{yerokhin:05:se})
\begin{align}\label{eq:subtr2}
\Delta E_{\rm subtr, YPS}^{(2+)} &\, =
 4\pi i\alpha \int_{C_H}\!\!d\omega \,
  \int_0^{\infty}\!\! dx_1dx_2\,\int_{-1}^{1}\!\!d\xi\,(x_1x_2)^2\,
  \Big\{
  {\cal F}_1\Big[ g_{a}(x_1)f_{a}(x_2)(x_1P_{\overline{l}_a}-x_2P_{l_a})
   + f_{a}(x_1)g_{a}(x_2)(x_2P_{\overline{l}_a}-x_1P_{l_a})\Big]
\nonumber \\ &
 +{\cal F}_2\Big[ g_{a}(x_1)g_{a}(x_2)P_{l_a}
    - f_{a}(x_1)f_{a}(x_2)P_{\overline{l}_a}\Big]
 +{\cal F}_3\Big[ g_{a}(x_1)g_{a}(x_2)P_{l_a}
    + f_{a}(x_1)f_{a}(x_2)P_{\overline{l}_a}\Big]
\Big\}\,,
\end{align}
where $P_l = P_l(\xi)$ are the Legendre polynomials, $l_a = |\kappa_a+1/2|-1/2$ and
$\overline{l}_a = |\kappa_a-1/2|-1/2$.
The remaining four-dimensional integration is carried out numerically.
In order to simplify the numerical integration, we make the following change of variables
\begin{align}\label{eq:subtr3}
  \int_0^{\infty}\!\! dx_1dx_2\,\int_{-1}^{1}\!\!d\xi\,(x_1x_2)^2\,F(x_1,x_2,\xi)
  =
  \int_0^{\infty}\!\! dy\, \int_0^1dr\,
    \int_{x_{\rm min}}^{x_{\rm max}}\!\!dx\,
        \frac{ry^3x}{a^4}\,\Big[F(x_1,x_2,\xi)+F(x_2,x_1,\xi)\Big]\,,
\end{align}
where $r = x_2/x_1$, $y = a\,x_1$, $a = 2 \sqrt{1-\vare_a^2}$,
$x^2 = x_1^2+x_2^2-2x_1x_2\xi$,
$x_{\rm min} = x_1-x_2$, and $x_{\rm max} = x_1+x_2$.

\end{widetext}

\subsection{SC scheme}

The approach introduced in Ref.~\cite{sapirstein:23} and referred to as the SC scheme
used the following approximation $G^{(2+)}_{a,\rm SC}$ to the many-potential Green function $G^{(2+)}$,
\begin{align}
    G_{a,\rm SC}^{(2+)}(\vare,\bfx_1,\bfx_2)  =
 \frac12\, V(x_1)\,\frac{\partial^2}{\partial \vare^2}\, G^{(0)}(\vare,\bfx_1,\bfx_2)\,V(x_2)\,.
\end{align}
Again, one subtracts and re-adds $G_{a,\rm SC}^{(2+)}$ in the many-potential contribution,
separating it into two parts,
\begin{equation}
    \Delta E_{\rm SE}^{(2+)} = \Delta E_{\rm acc, SC}^{(2+)} + \Delta E_{\rm subtr, SC}^{(2+)}\,.
\end{equation}
The first term $\Delta E_{\rm acc, SC}^{(2+)}$ is obtained from Eq.~(\ref{many5}) by applying the substitution
$G^{(2+)} \to G^{(2+)}- G_{a,\rm SC}^{(2+)}$ everywhere, whereas the subtraction term is easily
represented in the momentum space as
\begin{align}
\Delta E_{\rm subtr, SC}^{(2+)} =
    \frac12 \int \frac{d^3p}{(2\pi)^3}\,
        \overline{\psi}_{Va}(\bfp)\, \frac{\partial^2 \Sigma^{(0)}_R(\vare_a,\bfp)}{\partial \vare_a^2}\, \psi_{Va}(\bfp) \,,
\end{align}
where $\psi_{Va}(\bfp)$ is defined by Eq.~(\ref{wfVa}) and the derivative of $\Sigma^{(0)}_R(\vare_a,\bfp)$
in the Coulomb gauge
can be easily worked out from Eq.~(\ref{sefreecoul}). The numerical evaluation of the subtraction term
is fully analogous to that of the zero-potential term and does not cause any problems.

\section{ Numerical aspects}
\label{sec:3}

We now provide details of our numerical computations. In the present work we
assume the binding potential to be
the point-nucleus Coulomb potential. The advantage of this choice is
that the reference-state wave functions and the Dirac-Coulomb Green function
are known analytically. The computation of the point-nucleus
bound wave functions in coordinate and momentum space is described in
Ref.~\cite{mohr:00:rmp}. The radial Dirac-Coulomb Green function is represented
analytically in terms of regular and irregular Whittaker functions
\cite{mohr:00:rmp}. Our computations of the Whittaker functions
follow the approach described in Ref.~\cite{yerokhin:99:pra}.
In the present work we upgraded the algorithms from
Ref.~\cite{yerokhin:99:pra} to the quadruple arithmetics
(appr. 32 significant digits)
to address significant numerical cancellations arising
from subtracting the zero- and one-potential Green functions in the
definition of $G^{(2+)}$ for low-$Z$ ions, especially for hydrogen.

The calculation of the zero-potential term is quite straightforward.
Numerical integrations over the Feynman parameters present in the Coulomb
gauge require some care because of integrable singularities that may
appear around $x = 0$, $u = 0$. They are dealt with by breaking the
integration region into subintervals that become increasingly smaller
when approaching the singularities, e.g., $x_i = 0.001$, 0.01, 0.1,
and applying the Gauss-Legendre quadratures for each subinterval.

The computation of the one-potential term is more time consuming than that
of the zero-potential term, due to the larger number of integrations.
However, after the separation into the diagonal and nondiagonal in $p$
contributions in Eq.~(\ref{eq:6}) and the change of variables
in the nondiagonal part given by Eq.~(\ref{eq:7}), the calculation of the one-potential term becomes
relatively straightforward. The numerical integrations are performed
with the Gauss-Legendre quadratures. The integration
of the Feynman parameter $x$ has integrable singularities both around $x = 0$
and $x = 1$, which were dealt with in the same way as for the zero-potential
contribution.

The calculation of the many-potential contribution requires the integration contour
$C_{LH}$, with its parameters selected to optimize the numerical performance.
The main parameter $\Delta$, which separates the
low-and the high-energy parts, was chosen within the interval
$(3(\vare_a-\vare_{1s})/2,\Za\,\vare_a)$. For low $Z$, we usually use the
upper boundary, $\Delta = \Za\,\vare_a$. For high $Z$ and the accelerated
YPS scheme, we typically decrease the parameter $\Delta$ to make the partial-wave expansion
of the low-energy part converge faster. The parameter $\delta_{x,1}$ is fixed by
$\delta_{x,1} = \vare_a - \vare_{1s}$.
The choice of the parameter $\delta_y$ is a more nuanced matter. On one hand,
$\delta_y$ should not be too small, as it keeps the integration contour away from
the poles on the real axis, making the integrand smooth enough for
high-precision numerical integration. On the other hand, $\delta_y$ should not
be too large, as this could introduce numerical instabilities, given that the
photon propagator on the upper side of the integration contour in the lower $\omega$
half-plane is an exponentially  growing function.
The compromise is empirically found to be
around $\delta_y = (\vare_a - \vare_{1s})/2$. The exact choice of the
parameter $\delta_{x,2}$ does not matter, it was usually taken as
$\delta_{x,2} = 2\delta_{x,1}$.

The many-potential term can be split into two parts,
the instantaneous contribution (induced by the first term with $g_J(0)$ in brackets in Eq.~(\ref{many5}))
and the retarded contribution (which is the remainder
of Eq.~(\ref{many5})). The computation of the instantaneous part is
more complicated and we performed it separately.
The complications are twofold.
First, unlike the retarded part
$g_J(\omega)$, the instantaneous part of the photon propagator
$g_J(0)$ does not decay for large imaginary $\omega$.
Consequently, the accuracy of the integrand
for large imaginary $\omega$ is very sensitive to the accuracy of
the two-potential Green function, which suffers from numerical cancellations in this region.
Because of this, we had to use the quadruple arithmetics for its evaluation.

Second, the partial-wave expansion terms of the
instantaneous part show large cancellations between different regions of $\omega$
integration. To compensate this, the numerical $\omega$ integration needs to be
performed to a very high accuracy, even for high partial waves which
yield very small overall contributions.
Due to the presence of the instantaneous contribution, computations in the Coulomb gauge are
generally more complicated than in the Feynman gauge. By contrast, the computation of the
retarded contribution is very much analogous to that in the Feynman gauge described
in Ref.~\cite{yerokhin:05:se}.

We now turn to the evaluation of the subtraction term in the YPS scheme given by
Eq.~(\ref{eq:subtr2}). After the change of variables of Eq.~(\ref{eq:subtr3}),
the numerical integrations over the $\omega$, $y$, and $r$ variables are similar to that
in the many-potential contribution. The integration over $x$ is similar to the
$q$ integration in the one-potential term. However, the integrable singularity at
small values of $x$ is stronger here and requires some care. We split the
integration region into several subintervals, typically, with $x = 0.01x_1$,
$0.1x_1$, and apply Gauss-Legendre quadratures.

Finally, we address the extrapolation of the PW expansion, an
important component of the numerical approach, as it typically determines the
numerical uncertainty of the final result.
We perform explicit calculations of the PW terms up to
$|\kappa_{\rm max}| = 35$-40, then fit several last terms
(typically, 4-5) to polynomials in $1/|\kappa|$ and use the
results of the least-square fit to extrapolate the PW sum to infinity. We usually use
several fitting functions of the form
\begin{align}
\delta E_{|\kappa|} = \sum_{l = l_0}^{l_1} \frac{c_l}{|\kappa|^l}\,,
\end{align}
with free coefficients $c_l$ and different sets of $(l_0,l_1)$.
Typically, $(l_0,l_1) = (3,4)$, $(3,5)$, and $(3,6)$ were used.
The convergence of each extrapolation is analyzed and the best fitting function
is selected. The uncertainty is estimated on the basis of the change of the
best extrapolation result when varying the $|\kappa_{\rm max}|$ parameter
by 20\%.

It should be noted that the stability of the extrapolation heavily depends
on the numerical accuracy of the PW expansion terms.
While higher-order PW terms are typically very small and their numerical uncertainties contribute
negligibly to the error of the unextrapolated sum, they
can significantly affect the extrapolated tail of the
PW expansion. Generally, the more terms are included in the fitting function,
the greater the numerical accuracy of the PW terms has to be to achieve
a stable extrapolation pattern.

\section{ Numerical results}
\label{sec:4}

Numerical results for the one-loop self-energy correction are conveniently expressed in terms of the
dimensionless function $F_{\rm SE}(\Za)$,
\begin{equation} \label{Falpha}
\Delta E_{\rm SE} = mc^2\,\frac{\alpha}{\pi}\, \frac{(\Za)^4}{n^3}\, F_{\rm SE}(\Za) \,.
\end{equation}
The $\Za$ expansion of the function $F_{\rm SE}(\Za)$ has the form
\begin{align} \label{eq:se:za}
F_{\rm SE }(\Za) &\ = L\,A_{41} + A_{40} + (\Za)\,A_{50}
  \nonumber \\ &
+ (\Za)^2\,\bigl[L^2\,A_{62} + L\,A_{61}+ A_{60}\bigr]
  \nonumber \\ &
+ (\Za)^3 \bigl[L\,A_{71}+ A_{70}\bigr]
+ \ldots\,,
\end{align}
where $L = \ln(\Za)^{-2}$.
The analytical results for the coefficients $A_{41}$--$A_{61}$ are summarized in Ref.~\cite{yerokhin:18:hydr}.
The coefficient $A_{60}$ was calculated for different states and tabulated
in Refs.~\cite{pachucki:93,jentschura:96,bigot:03},
whereas the coefficient $A_{71} = \pi (139/64-\ln2)\delta_{l,0}$ is taken from
Ref.~\cite{karshenboim:97}. In order to assign an uncertainty to predictions based on
the $\Za$ expansion, we used the estimate for the unknown $A_{70}$ coefficient of $A_{70} = 0\pm 8\,A_{60}$,
where the numerical coefficient 8 was selected after comparing the $\Za$-expansion and all-order results
for the hydrogen $2P$ states.

We start by comparing our calculations of the one-loop self-energy correction
performed within the standard
potential-expansion approach in three different gauges. The first one is the Feynman gauge, which is technically
the simplest choice and is usually used in the literature.
The analysis of Snyderman in Ref.~\cite{snyderman:91} showed that the
individual terms in Feynman gauge contain spurious contributions of order $\alpha(\Za)^2$
which
cause numerical cancelations for low $Z$.
It was suggested in Ref.~\cite{snyderman:91} to use the Fried-Yennie (FY) gauge, in which
spurious terms are absent. Following this suggestion, we take
the FY gauge as our second choice.
Our third choice is the Coulomb gauge. Similarly to the FY gauge, the Coulomb
gauge is known for its soft infrared properties. Explicit numerical calculations
performed in Ref.~\cite{hedendahl:12} confirmed the absence of spurious contributions in
the one-loop self-energy in this gauge.

Table~\ref{tab:Z10} presents our numerical results obtained within these three gauges
for the $1s$ state of H-like neon ($Z = 10$). We observe significant cancelations between
the zero-, one- and many-potential terms in the Feynman gauge. These cancelations
are mostly absent in the FY gauge. However,
the PW expansion in the FY gauge converges (slightly) more slowly than in the Feynman gauge.
Since the convergence of the PW expansion is typically the limiting factor for the numerical
accuracy of practical calculations \cite{yerokhin:99:pra}, we conclude that the FY gauge does
not provide computational advantages over the Feynman gauge. Nevertheless, it could still
prove valuable for validating the numerical approach and the PW extrapolation procedure.

By contrast, numerical results obtained in the Coulomb gauge show both absence of
spurious cancelations and some (although, small) improvement in the convergence of the PW expansion.
The final results obtained in the three gauges are consistent with each other but have
different uncertainties, with the best accuracy achieved in the Coulomb gauge.
We conclude that the Coulomb gauge is the optimal choice of the gauge for calculations
in the low-$Z$ region. We should note, however, that calculations in the Coulomb gauge
are technically more complicated than in Feynman gauge, so the improvement comes
at a price.

Table~\ref{tab:Z10} presents also our numerical results obtained in two approaches with the accelerated
PW convergence, in the Coulomb gauge. We observe that in both cases the convergence of the PW
expansion is greatly accelerated, which leads to the improvement of accuracy of the final values
by 3-4 orders of magnitude.
The convergence of the YPS scheme is somewhat better than in the SC case and its expansion
terms do not change their sign. This leads to a more accurate result obtained in the YPS approach.

We now turn to the most difficult case, namely, hydrogen ($Z = 1$). So far accurate direct calculations for
the electron self-energy at $Z = 1$ were
reported only by Jentschura and Mohr \cite{jentschura:99:prl,jentschura:01:pra,jentschura:04:se}.
Table~\ref{tab:Z1} presents the breakdown of our numerical calculations in the YPS and SC accelerated-convergence
approaches for hydrogen. For the YPS scheme, we present results both in Feynman and in Coulomb gauge.
We observe large numerical cancelations in the Feynman gauge between the zero-, one-, and many-potential
contributions. The advantages of the Coulomb gauge in this case are evident.

In addition to the absence
of spurious contributions,
the PW expansion in the Coulomb gauge exhibits a significantly
(about an order of magnitude!)
better convergence compared to the Feynman gauge.
The performance of the two schemes is very similar in this case, apparently because
terms with three and more Coulomb interactions accounted for by the YPS approach are
not significant for such low $Z$. The YPS result is slightly more accurate than the SC one,
because the SC expansion terms probably change their sign for large $\kappa$ which makes the
extrapolation less stable.
The results of both approaches are consistent and agree well with the more precise value obtained
by Jentschura and Mohr.

An important advantage of the numerical approach of this work is that it can be applied for any $Z$,
including the case of $Z=1$, and not
only for the ground but also for highly excited states. Table~\ref{tab:tot} presents our
results obtained for all $n\le 5$ states of hydrogen and H-like ions with $Z = 5$, 10, 20, and 40, in
comparison with literature values.
The listed values are obtained with the YPS approach in the Coulomb gauge, for the
point nuclear model and the fine-structure fixed by $\alpha^{-1} = 137.036$, to be
directly comparable with the numerical results of Mohr and collaborators.

For the $D_{5/2}$, $F$, and $G$ states, our results are the first direct calculations
of the electron self-energy for $Z < 60$. (Le Bigot et al. \cite{bigot:01:se} reported
accurate results for these
states but for $Z\ge 60$ only.) For $Z = 1$ and 5, our calculations merely cross-check
the $\Za$-expansion predictions, which are remarkably accurate for these excited states,
but already starting with $Z=10$ we can identify contributions due to unknown terms of order
$\alpha (\Za)^7$ and higher.

For $nS$, $nP$, $nD_{3/2}$ states with $n = 3-5$ and $Z\ge 10$, our results
in Table~\ref{tab:tot} are compared with
those by Mohr and Kim \cite{mohr:92:a}. Excellent agreement is observed in all cases,
with our results typically providing 1–3 additional digits of precision.
For $n \le 2$ and $Z > 5$, we compare our results with the most accurate
previous calculations \cite{mohr:92:b,indelicato:98}.
Overall, the results are in good agreement, with only minor discrepancies observed for the
the $1S$ and $2S$ states at $Z = 10$ with results of
Indelicato and Mohr \cite{indelicato:98}.

Calculations by Jentschura and Mohr \cite{jentschura:99:prl,jentschura:01:pra,jentschura:04:se}
for $Z\le 5$ provided extremely precise results, surpassing the accuracy achieved in the present study.
However, these calculations are far more computationally expensive than those presented here
and have been conducted only for a limited set of states ($nS$ with $n \leq 4$ and $2P_j$) for ions with $Z = 1\ldots5$.
The comparison shown in Table~\ref{tab:tot} provides an important confirmation of these high-precision results.

Of particular importance is hydrogen,
whose theory is the basis for the
determination of the Rydberg constant \cite{tiesinga:21:codata18}.
The current theoretical uncertainty of the $1S$-$2S$ transition frequency in hydrogen is about 1~kHz \cite{tiesinga:21:codata18,yerokhin:24:tobe}, which corresponds to a fractional accuracy of $1 \times 10^{-7}$ for the $1S$ self-energy correction and $9 \times 10^{-7}$ for the $2S$ state. Until now, the calculation by Jentschura and Mohr \cite{jentschura:99:prl,jentschura:01:pra} was the only one to achieve this level of numerical accuracy for $Z = 1$. Our results thus provide the first independent cross-check of their results.

The extended tables of our numerical results are available in Supplementary Material \cite{suppl}.

\section{Summary}

We performed calculations of the one-loop electron self-energy in both the general
covariant gauge and the Coulomb gauge. Our results demonstrated that usage of the Coulomb
gauge in self-energy calculations may offer significant advantages,
especially for low nuclear charges.
These advantages are twofold. First, spurious
contributions leading to large numerical cancellations in the low-$Z$ region are absent.
Second, the convergence of the partial-wave expansion becomes faster.
When combined with the accelerated-convergence methods presented in
Refs.~\cite{yerokhin:05:se,sapirstein:23}, usage of the Coulomb gauge enabled us
to develop a highly efficient numerical approach
capable of producing accurate results for the self-energy corrections for any nuclear charge $Z$,
including $Z = 1$, and arbitrary excited states.
As a result, we extended the region of the self-energy calculations
for the $D_{5/2}$, $F$, and $G$ excited states as compared
with previously reported calculations, improved the
numerical accuracy of the literature results
for some excited states and values of $Z$, and performed an extensive independent
cross-check of previous numerical and $\Za$-expansion calculations.
We believe that the developed method will be useful in calculations
of higher-order self-energy corrections in the low-$Z$ region.


%
%
%

\appendix

\begin{widetext}

\section{Photon propagator}
\label{app:phot}

The photon propagator has the simplest form in the Feynman (F) gauge, where it is given by
\begin{align} \label{ph0}
D^{\mu\nu}_{F}(\omega,\bfx_1,\bfx_2) =&\ -\int \frac{d^3k}{(2\pi)^3}\,
  e^{i \bfk \cdot (\bfx_1-\bfx_2)}\,
    \frac{g^{\mu\nu}}{\omega^2 - \bfk^2+i0}
    =
    g^{\mu\nu}\, \frac{e^{i\sqrt{\omega^2+i0}\,x_{12}}}{4\pi x_{12}}
    \equiv
    g^{\mu\nu}\, D(\omega,x_{12})\,,
\end{align}
where $x_{12} = |\bfx_1-\bfx_2|$.
In the general covariant gauge the photon propagator is written as
\begin{align} \label{ph1}
D^{\mu\nu}(\omega,\bfx_1,\bfx_2)
\equiv D^{\mu\nu}_F(\omega,\bfx_1,\bfx_2) + \xi\,D^{\mu\nu}_{\xi}(\omega,\bfx_1,\bfx_2) =
-\int \frac{d^3k}{(2\pi)^3}\,
  e^{i \bfk \cdot (\bfx_1-\bfx_2)}\,
   \bigg[
    \frac{g^{\mu\nu}}{\omega^2 - \bfk^2+i0} + \xi \frac{k^{\mu}k^{\nu}}{(\omega^2 - \bfk^2+i0)^2}\Big|_{k^0 = \omega}
    \bigg]
    \,,
\end{align}
where $\xi$ is the gauge parameter.
The particular case of the general covariant gauge is the so-called Fried-Yennie gauge, with the gauge parameter $\xi = 2$.

In atomic physics, the photon propagator typically appears in combination with Dirac matrices, as
$\alpha_{\mu}\alpha_{\nu}D^{\mu\nu}$. This combination can be written in coordinate space as
\begin{align}\label{ph2}
\alpha_{1\mu}\alpha_{2\nu}D^{\mu\nu}(\omega,\bfx_1,\bfx_2) =&\
\bigg[
 1-\balpha_{1}\cdot\balpha_2  + \xi \big( \omega + i\balpha_1 \cdot \bnabla_1\big)
 \big( \omega - i\balpha_2 \cdot \bnabla_2\big)
 \left( -\frac1{2\omega}\frac{\partial}{\partial \omega}\right)
 \bigg]
 D(\omega,x_{12})\,.
\end{align}

The photon propagator in the Coulomb (C) gauge is given by
\begin{align}\label{ph3a}
D^{00}_{C}(\omega,\bfx_1,\bfx_2) = \frac1{4\pi x_{12}}\,,&\  \ \ \ D^{0i}_{C}(\omega,\bfx_1,\bfx_2) = D^{i0}_{C}(\omega,\bfx_1,\bfx_2) = 0\,,\\
D^{ij}_{C}(\omega,\bfx_1,\bfx_2) = &\
\int \frac{d^3k}{(2\pi)^3}\,
  \frac{e^{i \bfk \cdot (\bfx_1-\bfx_2)}}{\omega^2 - \bfk^2+i0}\,
   \bigg[
    \delta^{ij}- \frac{k^{i}k^{j}}{\bfk^2}
    \bigg]
    \,,
\end{align}
where $i$, $j = 1, 2, 3$. The combination $\alpha_{\mu}\alpha_{\nu}D^{\mu\nu}_C$
is can be written as
\begin{align}\label{ph3}
\alpha_{1\mu}\alpha_{2\nu}D^{\mu\nu}_{C}(\omega,\bfx_1,\bfx_2) =&\
D(0,x_{12})
 -\balpha_{1}\cdot\balpha_2\,
 D(\omega,x_{12})
 + \frac1{\omega^2}\,\balpha_1 \cdot \bnabla_1
 \,\balpha_2 \cdot \bnabla_2
 \Big[
 D(\omega,x_{12})-D(0,x_{12})
 \Big]\,.
\end{align}
It is sometimes convenient to transform the above expression further, to take the following form
\begin{align}\label{ph4}
\alpha_{1\mu}\alpha_{2\nu}D^{\mu\nu}_{C}(\omega,\bfx_1,\bfx_2) =&\
D(0,x_{12})
 -\balpha_1\cdot\balpha_2\, D_1(\omega,x_{12})
 + (\balpha_1\cdot\hat{\bfx}_{12}) (\balpha_2\cdot\hat{\bfx}_{12})\,  D_2(\omega,x_{12})
 \,,
\end{align}
where
\begin{align}\label{ph5}
D_1(\omega,x) =&\ \frac1{4\pi\omega^2x^3}\,\Big[ 1-(1-i\omega x - \omega^2x^2)e^{i\sqrt{\omega^2+i0}\, x}\Big]\,,
 \nonumber \\
D_2(\omega,x) =&\  \frac1{4\pi\omega^2x^3}\,\Big[3- (3-3i\omega x - \omega^2x^2)e^{i\sqrt{\omega^2+i0}\, x}\Big]\,.
\end{align}

\section{Free self-energy operator in general covariant gauge}
\label{sec:freese1}

The free-electron self-energy operator in the general covariant gauge
can be evaluated in the same way as in the Feynman gauge,
see Ref.~\cite{yerokhin:99:pra}. The result is
\begin{equation}
\Sigma^{(0)}_{R}(\rp) = \frac{\alpha}{4\pi}\,\Big[a(\rho) + \slashed{p}\,b(\rho)\Big]\,,
\end{equation}
where $\rp$ is the four-vector $\rp = (p_0,\bfp)$, $\slashed{p} = p^{\mu}\gamma_{\mu}$
and the scalar functions $a$ and $b$ are given by
\begin{align}
a(\rho) =  2(1+\xi) + \frac{(4+\xi)\rho}{1-\rho} \ln\rho\,, \ \ \
b(\rho) =  -\frac{2-\rho}{1-\rho}\, (1+\xi)\,\Big( 1 + \frac{\rho}{1-\rho}\ln \rho\Big)\,,
\end{align}
with $\rho = 1 - \rp^2 = 1- p_0^2+\bfp\,^2$.
In the Feynman gauge ($\xi = 0$), the above expressions coincide with Eqs.~(A5)-(A7) of
Ref.~\cite{yerokhin:99:pra}.

\section{Free self-energy operator in Coulomb gauge}
\label{sec:freese2}

The renormalized free-electron self-energy operator in the Coulomb gauge was derived by Adkins
\cite{adkins:86}.
It is expressed as
\begin{align}\label{sefreecoul}
\Sigma^{(0)}_{R,\rm Coul}(\rp) =&\  \frac{\alpha}{4\pi}
 \int_0^1 dx\, du\,
  \Big[
  	a + p_0\gamma^0\,b + \bgamma\cdot\bfp\, c
 \Big]\,,
\end{align}
 where
\begin{align}
a = &\  -\frac1{\sqrt{x}}\,\ln {\cal X} - 2\ln {\cal Y}\,, \ \ \ \
b = 2(1-x) \, \ln {\cal Y} - \frac12 \,, \ \ \ \
c =  \frac{19}{6} - \frac{1-x}{\sqrt{x}}\, \ln {\cal X} - 2(1-x)\, \ln {\cal Y} + 2\sqrt{x}\,\ln {\cal Z}\,,
\end{align}
with ${\cal X} = 1 + \bfp\,^2(1-x)$, ${\cal Y} = 1- (p_0^2-\bfp\,^2)(1-x)$, and ${\cal Z} = 1- p_0^2 (1-u)+ \bfp\,^2(1-xu)$.
The integration over $u$ can easily be carried out analytically, but we prefer to keep the expression in its
present form because its numerical evaluation is very cheap.

\section{Free-electron vertex operator in general covariant gauge}
\label{sec:freever1}

The free-electron vertex operator in the general covariant gauge can be represented
as a sum of the Feynman-gauge and the gauge-dependent parts,
\begin{align}
\Gamma^{\mu}(\rp^{\prime},\rp) =  \Gamma^{\mu}_F(\rp^{\prime},\rp) + \xi \Gamma^{\mu}_{\xi}(\rp^{\prime},\rp)\,.
\end{align}
The Feynman-gauge part is given in Appendix B of Ref.~\cite{yerokhin:99:pra}, so we
concentrate here on the gauge-dependent part.
The time component of the gauge-dependent part of the renormalized free-electron vertex operator is expressed as
\begin{align}
\Gamma^{\,0}_{\xi,R}(\rp^{\prime},\rp) = \frac{\alpha}{4\pi}
\Big[ \gamma^0A + \slashed{p}^{\prime}\,B + \slashed{p}\,C
 + \slashed{p}^{\prime}\gamma^0\,E+ \gamma^0\slashed{p}\,F
 + \slashed{p}^{\prime}\slashed{p}\,G + H
 \Big]\,,
\end{align}
where, for $p^{\prime}_0 = p_0 = \vare$,
\begin{align}
A =&\ (\rp^{\prime2}+m^2)C_{11} + (\rp^2+m^2)C_{12} + C_{24}\,, \nonumber \\
B =&\ 2\vare\big[ (\rp^{\prime2}+m^2)D_{21} + (\rp^2+m^2)D_{23}+C_{11}\big]\,, \nonumber \\
C =&\ 2\vare\big[ (\rp^{2}+m^2)D_{22} + (\rp^{\prime2}+m^2)D_{23}+C_{12}\big]\,, \nonumber \\
E =&\  m\big[(-\rp^{\prime2}+2\rp^{\prime}\cdot\rp)D_{21} + \rp^2D_{22}+ 2\rp^2 D_{23}]\,, \nonumber \\
F =&\  m\big[ \rp^{\prime2}D_{21}+ (-\rp^{2}+2\rp^{\prime}\cdot\rp)D_{22} +2\rp^{\prime2} D_{23}]\,, \nonumber \\
G =&\ -2m\vare \big[D_{21}+D_{22}+2D_{23}\big]\,, \nonumber \\
H =&\ 2m\vare \big[ C_0 - \rqq^2D_{23}+C_{11}+C_{12}\big]\,.
\end{align}
Here,
\begin{eqnarray} \label{vertex3}
\left( \begin{array}{c} D_{21} \\ D_{22} \\D_{23} \end{array} \right)
        &=&
        \int_0^1dy\, \frac{1}{(y\rp^{\prime}+(1-y)\rp)^4}
\left( \begin{array}{c} y^2 \\ (1-y)^2 \\y(1-y) \end{array} \right)
        [-2 + (1+2Y)\ln X] \,,
\end{eqnarray}
and all other notations are from Appendix B of Ref.~\cite{yerokhin:99:pra}.

\section{Free-electron vertex operator in Coulomb gauge}
\label{sec:freever2}

The renormalized free-electron vertex operator in the Coulomb gauge was derived by Adkins
\cite{adkins:86}.
For the purposes of the present investigation, we need only the time component of the vertex
operator, which is expressed as
\begin{align}
\Gamma^{\,0}_{R,\rm Coul}(\rp^{\prime},\rp) =&\
 \frac{\alpha}{4\pi}
  \int_0^1 dx\,ds\,du\,
   \Big[
   \gamma^0{\cal A}
   + \bgamma\cdot\bfp\,^{\prime}\,{\cal B}
   + \bgamma\cdot\bfp\,\,{\cal C}
   + \bgamma\cdot\bfp\,^{\prime} \gamma^0 \bgamma\cdot\bfp\,\,{\cal D}
   + \bgamma\cdot\bfp\,^{\prime}\gamma^0\,{\cal E}
   + \gamma^0\bgamma\cdot\bfp\,\,{\cal F}
   + {\cal H}
   \Big]\,,
\end{align}
where
\begin{align}
{\cal A} = &\ \frac{1}{\sqrt{x}\Delta_X}
 \left\{ 2xu(1-u)\bfk\,^2 + (1-x)[u\bfp\,^{\prime2}+(1-u)\bfp\,^2]  \right\}
 +  \frac2{\Delta_Y} \bigg\{
 1-x^2 + (xq_0-p_0')(xq_0-p_0) + x^2 u(1-u)k_0^2
 \nonumber \\ &
 -\frac{x}2(3-2x) \left[ u(1-p_0^{\prime2})+ (1-u)(1-p_0^2)\right]
 + \bfk^2 \left[ -1 + \frac{x}2 - 2x^2u(1-u)\right]
 + \bfp\,^{\prime2} \left[ 1-\frac{x}2 - \frac{x}2(5-4x)u\right]
 \nonumber \\ &
 + \bfp\,^2 \left[ 1- \frac{x}2 - \frac{x}2 (5-4x)(1-u)\right]
 \bigg\}
+ \frac{\sqrt{x}}{\Delta_Z}(-2\bfp\,'\cdot\bfp)
+ \frac{2x\sqrt{x}s}{\Delta_Z^2}(-\bfp\,'\cdot\bfp) \,\bfq\,^2
- \ln \Delta
\,,  \\
{\cal B} = &\ \frac{1}{\sqrt{x}\Delta_X}k_0 u(-1-x+2xu) + \frac2{\Delta_Y}(-u)x(2xq_0-p'_0-p_0)
   + \frac{\sqrt{x}}{\Delta_Z}(sq_0-p_0)
\nonumber \\ &
   + \frac{2x\sqrt{x}s}{\Delta_Z^2}u [\bfp\,' \cdot \bfq\, (sq_0-p_0) + \bfp\, \cdot \bfq\, (sq_0-p_0')]\,,
   \\
{\cal C} = &\ \frac{1}{\sqrt{x}\Delta_X}k_0(1-u)(1-x+2xu) + \frac2{\Delta_Y}(u-1)x(2xq_0-p'_0-p_0)
   + \frac{\sqrt{x}}{\Delta_Z}(sq_0-p_0')
\nonumber \\ &
   + \frac{2x\sqrt{x}s}{\Delta_Z^2}(1-u) [\bfp\,' \cdot \bfq\, (sq_0-p_0) + \bfp\, \cdot \bfq\, (sq_0-p_0')]\,,
   \\
{\cal D} = &\ -\frac{1-x}{\sqrt{x}\Delta_X}  -\frac2{\Delta_Y} + \frac{2\sqrt{x}}{\Delta_Z}
   + \frac{2x\sqrt{x}s}{\Delta_Z^2}\bfq\,^2 \,, \\
{\cal E} = &\  -\frac{1}{\sqrt{x}\Delta_X} + \frac{\sqrt{x}}{\Delta_Z} + \frac{2x\sqrt{x}s}{\Delta_Z^2}(-u)\bfq\cdot\bfk\,, \\
{\cal F} = &\  -\frac{1}{\sqrt{x}\Delta_X} + \frac{\sqrt{x}}{\Delta_Z} + \frac{2x\sqrt{x}s}{\Delta_Z^2}(1-u)\bfq\cdot\bfk\,, \\
{\cal H} = &\ \frac{1}{\sqrt{x}\Delta_X}k_0(1-2u)
 +  \frac2{\Delta_Y}(2xq_0-p'_0-p_0)\,,
\end{align}
where $\rqq = u\rp'+(1-u)\rp$ and $\rk = \rp-\rp'$. Furthermore, $\Delta = 1 - u(1-u)(k_0^2- \bfk\,^2)\,,$
\begin{align}
\Delta_X = &\ 1 - u(1-u)(k_0^2-x  \bfk\,^2) + (1-x) [ u  \bfp\,^{\prime2} + (1-u) \bfp\,^{2}]\,,\\
\Delta_Y = &\ x - xu(1-u) (k_0^2-\bfk\,^2) + (1-x) u (1-p_0^{\prime2}+ \bfp\,^{\prime2}) + (1-x)(1 -u)(1-p_0^2+ \bfp\,^{2})\,,\\
\Delta_Z = &\ s - su(1-u) (k_0^2-x\bfk\,^2)
 + (1-s) u (1-p_0^{\prime2})  + (1-s) (1-u) (1-p_0^{2})
 + (1-xs) [ u  \bfp\,^{\prime2} + (1-u) \bfp\,^{2}]\,.
\end{align}
Integration over one of the Feynman parameters ($s$) in the above formulas can be easily carried out analytically
in terms of basic integrals $J_{01}$, $J_{11}$, $J_{12}$, $J_{22}$, where
\begin{align}
J_{ik} = \int_0^1ds\, \frac{s^i}{(1+sZ)^k}\,,
\end{align}
and $Z$ comes from representing $\Delta_Z = \Delta_{Z,s=0}\,(1 + sZ)$.

\section{Approximate many-potential Green function $\bm G_{\bm a}^{\bm{(2+)}}$}
\label{app:Ga}

Using the closed-form formula for the free Green function \cite{mohr:74:a}
\begin{eqnarray}
    G^{(0)}(E,\bfx_1,\bfx_2) &=& -\left[
\left(\frac{c}{x_{12}}+\frac1{x_{12}^2}\right)
           i\, \balpha \cdot \bfx_{12} +\beta +E \right]\,
                  \frac{e^{-c\,x_{12}}}{4\pi x_{12}}\,,
\end{eqnarray}
where $\balpha$ and $\beta$ are the Dirac matrices and $c = \sqrt{1-E^2}\,$, we express $G_a^{(2+)}$ in Eq.~(\ref{13}) as
\begin{align}
G_{a}^{(2+)}(E,\bfx_1,\bfx_2) = i\balpha\cdot\bfx_{12}\,G_1(E,\Omega,x_{12}) + \beta\,G_2(E,\Omega,x_{12}) + G_3(E,\Omega,x_{12})\,,
\end{align}
with
\begin{align}
G_1(E,\Omega,x) = &\ -\frac1{4\pi x}
 \bigg[ e^{-c^{\prime}x} \Big( \frac{c^{\prime}}{x}+ \frac1{x^2}\Big)
      - e^{-c\,x} \Big( \frac{c}{x}+ \frac1{x^2}+ \Omega E\Big)
      \bigg]
      \,, \nonumber \\
G_2(E,\Omega,x) = &\ -\frac1{4\pi x}
 \bigg[  e^{-c^{\prime}x}
      - e^{-c\,x} \Big( 1+\Omega\, \frac{E}{c}\,x\Big)
      \bigg]
      \,, \nonumber \\
G_3(E,\Omega,x) = &\ -\frac1{4\pi x}
 \bigg[  e^{-c^{\prime}x} (E+\Omega)
      - e^{-c\,x} \Big( E+ \Omega+\Omega\, \frac{E^2}{c}\, x\Big)
      \bigg]
      \,,
\end{align}
and $c^{\prime} = \sqrt{1-(E+\Omega)^2}\,$.
\end{widetext}

%
%
\begingroup
\begin{table*}[htb]
\begin{center}
\begin{minipage}{16.0cm}
\caption{Individual contributions to the one-loop self-energy
correction for the $1s$ state and $Z = 10$.
Results are presented
for the standard potential-expansion approach
in the Feynman (Feyn), Fried-Yennie (FY), and
Coulomb (Coul) gauges,
and for two accelerated-convergence approaches, Acc. YPS from Ref.~\cite{yerokhin:05:se} and
Acc. SC from Ref.~\cite{sapirstein:23}.
Units are $F(Z\alpha)$.
\label{tab:Z10}
}
\begin{ruledtabular}
\begin{tabular}{lw{5.8}w{5.8}w{5.8}w{7.13}w{5.13}}
                                &  \multicolumn{3}{c}{Standard}
				                                &  \multicolumn{1}{c}{Acc. YPS}
  				                                            &  \multicolumn{1}{c}{Acc. SC}
\\
\cline{2-4}
\\[-7pt]
                                &  \multicolumn{1}{c}{Feyn}
                                               &  \multicolumn{1}{c}{FY}
                                                                &  \multicolumn{1}{c}{Coul}
                                                                &  \multicolumn{1}{c}{Coul}
                                                                &  \multicolumn{1}{c}{Coul}
			 \\
\hline
\\[-7pt]
 Zero-pot.                            & -828.250\,01      & -1.266\,74    &  5.502\,20      &  5.502\,205\,84      &   5.502\,205\,84 \\
  One-pot.                            &  644.228\,50      &  6.693\,05    & -0.278\,26      & -0.278\,264\,37      &  -0.278\,264\,37 \\
 Subtraction                          &                   &              &                  & -0.485\,684\,60      &  -0.791\,764\,94 \\
 $|\kappa| =     $     1              &  183.505\,54      &  7.549\,46    & -0.262\,57      & -0.023\,706\,24      &   0.222\,516\,30 \\
                       2              &    3.339\,31      & -4.713\,07    & -0.145\,41      & -0.062\,867\,81      &  -0.001\,734\,82 \\
                       3              &    0.863\,90      & -1.381\,76    & -0.040\,35      &  0.001\,006\,80      &   0.001\,294\,69 \\
                       4              &    0.367\,77      & -0.668\,08    & -0.025\,03      &  0.000\,729\,76      &   0.000\,278\,72 \\
                       5              &    0.193\,45      & -0.391\,13    & -0.017\,49      &  0.000\,332\,88      &   0.000\,037\,32 \\
                       6              &    0.114\,57      & -0.254\,12    & -0.012\,93      &  0.000\,163\,57      &  -0.000\,025\,88 \\
                       7              &    0.073\,33      & -0.176\,50    & -0.009\,90      &  0.000\,088\,09      &  -0.000\,040\,49 \\
                       8              &    0.049\,60      & -0.128\,49    & -0.007\,79      &  0.000\,051\,19      &  -0.000\,040\,59 \\
                       9              &    0.034\,99      & -0.096\,88    & -0.006\,26      &  0.000\,031\,62      &  -0.000\,036\,50 \\
                      10              &    0.025\,52      & -0.075\,08    & -0.005\,11      &  0.000\,020\,50      &  -0.000\,031\,61 \\
                      11              &    0.019\,12      & -0.059\,48    & -0.004\,23      &  0.000\,013\,83      &  -0.000\,027\,01 \\
                      12              &    0.014\,65      & -0.048\,00    & -0.003\,55      &  0.000\,009\,65      &  -0.000\,023\,00 \\
                      13              &    0.011\,44      & -0.039\,33    & -0.003\,00      &  0.000\,006\,92      &  -0.000\,019\,61 \\
                      14              &    0.009\,08      & -0.032\,65    & -0.002\,57      &  0.000\,005\,09      &  -0.000\,016\,78 \\
                      15              &    0.007\,32      & -0.027\,40    & -0.002\,21      &  0.000\,003\,81      &  -0.000\,014\,43 \\
16$\ldots$35                          &    0.038\,82      & -0.177\,80    & -0.015\,94      &  0.000\,014\,87\,(1) &  -0.000\,099\,87\,(5) \\
36$\ldots\infty$(extr.)               &    0.007\,13\,(35)& -0.052\,0\,(7)& -0.005\,51\,(18)&  0.000\,000\,93\,(3)&  -0.000\,030\,47\,(15) \\[2pt]
\hline\\[-7pt]
	 Total                            &    4.654\,01\,(35)&  4.653\,9\,(7)&  4.654\,08\,(18)&  4.654\,162\,33\,(3)&   4.654\,162\,50\,(16) \\
Ref.~\cite{indelicato:98}             &                   &               &                 &  4.654\,161\,9\,(1)\\
\end{tabular}
\end{ruledtabular}
\end{minipage}
\end{center}
\end{table*}
\endgroup

%
%
\begingroup
\begin{table*}[htb]
\begin{center}
\begin{minipage}{16.0cm}
\caption{Individual contributions to the one-loop self-energy
correction for the $1s$ state of hydrogen ($Z = 1$),
in the accelerated convergence scheme of Ref.~\cite{yerokhin:05:se}
(Acc. YPS) and of Ref.~\cite{sapirstein:23} (Acc. SC),
and Feynman or Coulomb gauge.
Units are $F(Z\alpha)$.
\label{tab:Z1}
}
\begin{ruledtabular}
\begin{tabular}{lw{5.12}w{5.12}w{5.12}}
				                                &  \multicolumn{2}{c}{Acc. YPS}
  				                                            &  \multicolumn{1}{c}{Acc. SC}
\\
\cline{2-3}
\\[-7pt]
                                                                &  \multicolumn{1}{c}{Feyn}
                                                                &  \multicolumn{1}{c}{Coul}
                                                                &  \multicolumn{1}{c}{Coul}
			 \\
\hline\\[-5pt]
 Zero-pot.                            &-168176.156\,251      & 13.849\,474\,1      & 13.849\,474\,1  \\
  One-pot.                            & 148579.466\,946      & -2.879\,681\,6      & -2.879\,681\,6  \\
 Subtr.                               &    216.681\,287\,(7) & -0.610\,070\,7\,(2) & -1.127\,787\,5  \\
 $|\kappa| =     $     1              &  19365.485\,663      &  0.043\,447\,0      &  0.475\,625\,7  \\
                       2              &     23.443\,958      & -0.091\,215\,7      & -0.004\,723\,6  \\
                       3              &      1.217\,848      &  0.002\,214\,0      &  0.001\,962\,3  \\
                       4              &      0.131\,074      &  0.001\,034\,7      &  0.000\,785\,7  \\
                       5              &      0.024\,727      &  0.000\,533\,5      &  0.000\,396\,6  \\
                       6              &      0.008\,445      &  0.000\,309\,4      &  0.000\,228\,1  \\
                       7              &      0.004\,243      &  0.000\,195\,0      &  0.000\,142\,8  \\
                       8              &      0.002\,535      &  0.000\,130\,4      &  0.000\,094\,8  \\
                       9              &      0.001\,643      &  0.000\,091\,2      &  0.000\,065\,8  \\
                      10              &      0.001\,120      &  0.000\,066\,1      &  0.000\,047\,2  \\
                      11              &      0.000\,791      &  0.000\,049\,3      &  0.000\,034\,8  \\
                      12              &      0.000\,575      &  0.000\,037\,6      &  0.000\,026\,3  \\
                      13              &      0.000\,428      &  0.000\,029\,3      &  0.000\,020\,2  \\
                      14              &      0.000\,326      &  0.000\,023\,1      &  0.000\,015\,8  \\
                      15              &      0.000\,252      &  0.000\,018\,6      &  0.000\,012\,5  \\
16$\ldots$35                          &      0.001\,096\,(3) &  0.000\,095\,4\,(6) &  0.000\,056\,5\,(6)  \\
36$\ldots\infty$(extr.)               &      0.000\,086\,(2) &  0.000\,012\,8\,(3) & -0.000\,002\,1\,(8)  \\[2pt]
\hline\\[-7pt]
	 Total                        &     10.316\,793\,(8) & 10.316\,793\,5\,(7) & 10.316\,794\,0\,(10)  \\
    Ref.~\cite{jentschura:01:pra}                        &                      & 10.316\,793\,650\,(1)\\
\end{tabular}
\end{ruledtabular}
\end{minipage}
\end{center}
\end{table*}
\endgroup

%
%
\begingroup
\begin{table*}[htb]
\begin{center}
\begin{minipage}{16.0cm}
\caption{The one-loop electron self-energy correction for hydrogen and hydrogen-like ions, in terms of $F(Z\alpha)$.
\label{tab:tot}
}
\begin{ruledtabular}
\begin{tabular}{lw{5.12}w{5.12}w{5.12}w{5.12}w{5.12}}
\multicolumn{1}{c}{State}
				                           &  \multicolumn{1}{c}{$Z = 1$}
    				                                &  \multicolumn{1}{c}{$Z = 5$}
    				                                &  \multicolumn{1}{c}{$Z = 10$}
    				                                &  \multicolumn{1}{c}{$Z = 20$}
    				                                &  \multicolumn{1}{c}{$Z = 40$}
\\[2pt]
\hline\\[-7pt]
$1S_{1/2}$  &  10.316\,793\,5\,(7)      &  6.251\,627\,05\,(4)    &  4.654\,162\,33\,(3)  &  3.246\,255\,62\,(2)   &  2.135\,228\,44\,(2)  \\
            &  10.316\,793\,650\,(1)^a  &  6.251\,627\,078\,(1)^a &  4.654\,161\,9\,(1)^c &  3.246\,255\,7\,(4)^c  &  2.135\,228\,4\,(1)^e\\[2pt]
$2S_{1/2}$  &  10.546\,825\,0\,(4)      &  6.484\,860\,2\,(4)     &  4.894\,416\,1\,(2)   &  3.506\,647\,70\,(5)   &  2.454\,829\,06\,(3)  \\
            &  10.546\,825\,185\,(5)^a  &  6.484\,860\,42\,(1)^a  &  4.894\,444\,4\,(6)^c &  3.506\,647\,8\,(2)^c  &  2.454\,829\,2\,(7)^e  \\[2pt]
$3S_{1/2}$  &  10.605\,614\,0\,(2)      &  6.543\,385\,8\,(2)     &  4.952\,410\,3\,(4)   &  3.563\,302\,3\,(2)    &  2.508\,273\,0\,(1)  \\
            &  10.605\,614\,22\,(5)^b   &  6.543\,385\,98\,(5)^b  &  4.952\,4\,(2)^d      &  3.563\,3\,(1)^d       &  2.508\,3\,(1)^d  \\[2pt]
$4S_{1/2}$  &  10.629\,388\,(2)         &  6.566\,758\,7\,(7)     &  4.974\,919\,5\,(8)   &  3.583\,402\,7\,(3)    &  2.521\,506\,2\,(2)  \\
            &  10.629\,388\,4\,(2)^b    &  6.566\,758\,8\,(2)^b   &  4.974\,9\,(4)^d      &  3.583\,4\,(1)^d       &  2.521\,5\,(1)^d  \\[2pt]
$5S_{1/2}$  &  10.641\,349\,(2)         &  6.578\,390\,(3)        &  4.985\,839\,(1)      &  3.592\,309\,0\,(4)    &  2.524\,604\,5\,(7)  \\
            &                           &                         &  4.985\,8\,(6)^d      &  3.592\,3\,(2)^d       &  2.524\,6\,(1)^d  \\[2pt]
$2P_{1/2}$  &  -0.126\,395\,(1)         & -0.122\,775\,(4)        & -0.114\,851\,(1)      & -0.092\,519\,2\,(6)    & -0.031\,049\,9\,(3 )  \\
            &  -0.126\,396\,37\,(1)^a   & -0.122\,774\,94\,(1)^a  & -0.114\,852\,(2)^c    & -0.092\,519\,0\,(1)^c  & -0.031\,050\,0\,(4)^e  \\[2pt]
$3P_{1/2}$  &  -0.115\,458\,1\,(8)      & -0.111\,259\,(2)        & -0.102\,047\,9\,(5)   & -0.076\,016\,1\,(5)    & -0.004\,189\,4\,(7)  \\
            &  -0.115\,461\,(4)^f       & -0.111\,4\,(4)^f        & -0.102\,1\,(2)^d      & -0.076\,0\,(1)^d       & -0.004\,1\,(1)^d  \\[2pt]
$4P_{1/2}$  &  -0.110\,424\,(2)         & -0.106\,015\,(6)        & -0.096\,340\,(5)      & -0.068\,975\,2\,(3)    &  0.006\,336\,6\,(6)  \\
            &  -0.110\,427\,(4)^f       & -0.106\,2\,(5)^f        & -0.096\,3\,(4)^d      & -0.069\,0\,(2)^d       &  0.006\,4\,(1)^d  \\[2pt]
$5P_{1/2}$  &  -0.107\,645\,(5)         & -0.103\,139\,(8)        & -0.093\,243\,(4)      & -0.065\,255\,2\,(2)    &  0.011\,561\,(1)  \\
            &  -0.107\,648\,(4)^f       & -0.103\,3\,(5)^f        & -0.093\,3\,(6)^d      & -0.065\,2\,(4)^d       &  0.011\,6\,(1)^d  \\[2pt]
$2P_{3/2}$  &   0.123\,498\,(2)         &  0.125\,623\,(2)        &  0.130\,354\,7\,(9)   &  0.143\,839\,1\,(4)    &  0.179\,594\,8\,(1)  \\
            &   0.123\,498\,56\,(1)^a   &  0.125\,623\,30(1)^a    &  0.130\,350\,7(3)^c   &  0.143\,838\,75\,(4)^c &  0.179\,594\,9\,(4)^e\\[2pt]
$3P_{3/2}$  &   0.134\,414\,(2)         &  0.136\,794\,(2)        &  0.142\,085\,9\,(3)   &  0.157\,185\,7\,(4)    &  0.197\,728\,2\,(6)  \\
            &   0.134\,413\,(2)^f       &  0.136\,7\,(2)^f        &  0.142\,1\,(2)^d      &  0.157\,2\,(1)^d       &  0.197\,7\,(1)^d  \\[2pt]
$4P_{3/2}$  &   0.139\,439\,(2)         &  0.141\,909\,(2)        &  0.147\,395\,(2)      &  0.163\,046\,0\,(2)    &  0.205\,168\,0\,(4)  \\
            &   0.139\,439\,(2)^f       &  0.141\,8\,(2)^f        &  0.147\,7\,(4)^d      &  0.163\,0\,(1)^d       &  0.205\,2\,(1)^d  \\[2pt]
$5P_{3/2}$  &   0.142\,215\,(5)         &  0.144\,724\,(3)        &  0.150\,297\,(1)      &  0.166\,190\,8\,(2)    &  0.208\,962\,2\,(4)  \\
            &   0.142\,215\,(2)^f       &  0.144\,6\,(3)^f        &  0.150\,2\,(6)^d      &  0.166\,2\,(1)^d       &  0.208\,9\,(1)^d  \\[2pt]
$3D_{3/2}$  &  -0.043\,019\,(2)         & -0.042\,927\,(1)        & -0.042\,708\,(3)      & -0.042\,020\,(2)       & -0.039\,617\,8\,(7)  \\
            &  -0.043\,0183\,3\,(2)^f   & -0.042\,929\,(2)^f      & -0.042\,8\,(2)^d      & -0.042\,0\,(1)^d       & -0.039\,6\,(1)^d  \\[2pt]
$4D_{3/2}$  &  -0.041\,007\,(3)         & -0.040\,903\,(3)        & -0.040\,661\,(4)      & -0.039\,880\,(1)       & -0.037\,056\,5\,(5)  \\
            &  -0.041\,0059\,6\,(2)^f   & -0.040\,907\,(2)^f      & -0.040\,3\,(4)^d      & -0.039\,9\,(1)^d       & -0.037\,1\,(1)^d  \\[2pt]
$5D_{3/2}$  &  -0.039\,861\,(4)         & -0.039\,752\,(3)        & -0.039\,496\,(4)      & -0.038\,662\,7\,(7)    & -0.035\,604\,4(5)  \\
            &  -0.039\,8592\,2\,(2)^f   & -0.039\,754\,(2)^f      & -0.039\,6\,(6)^d      & -0.038\,7\,(1)^d       & -0.035\,6\,(1)^d  \\[2pt]
$3D_{5/2}$  &   0.040\,317\,(1)         &  0.040\,432\,3\,(6)     &  0.040\,734\,(2)      &  0.041\,751\,(2)       &  0.045\,175\,4\,(5)  \\
            &   0.040\,316\,18\,(9)^f   &  0.040\,43\,(1)^f       &  0.040\,73\,(9)^f     &  0.041\,7\,(7)^f       &   \\[2pt]
$4D_{5/2}$  &   0.042\,329\,(1)         &  0.042\,460\,(4)        &  0.042\,801\,(2)      &  0.043\,959\,2\,(5)    &  0.047\,931\,7\,(5)  \\
            &   0.042\,328\,7\,(1)^f    &  0.042\,46\,(1)^f       &  0.042\,8\,(1)^f      &  0.043\,9\,(8)^f       &   \\[2pt]
$5D_{5/2}$  &   0.043\,476\,(2)         &  0.043\,612\,(6)        &  0.043\,973\,(2)      &  0.045\,196\,(2)       &  0.049\,415\,0\,(2)  \\
            &   0.043\,475\,6\,(1)^f    &  0.043\,61\,(1)^f       &  0.044\,0\,(1)^f      &  0.045\,1\,(8)^f       &   \\[2pt]
$4F_{5/2}$  &  -0.021\,498\,(2)         & -0.021\,480\,(4)        & -0.021\,438\,(4)      & -0.021\,309\,(3)       & -0.020\,920\,(2)  \\
            &  -0.021\,497\,019\,(7)^f  & -0.021\,480\,9\,(9)^f   & -0.021\,441\,(7)^f    & -0.021\,32\,(6)^f      &   \\[2pt]
$5F_{5/2}$  &  -0.020\,874\,(2)         & -0.020\,857\,(5)        & -0.020\,809\,(10)     & -0.020\,668\,(3)       & -0.020\,230\,3\,(7)  \\
            &  -0.020\,872\,240\,(7)^f  & -0.020\,854\,4\,(9)^f   & -0.020\,811\,(7)^f    & -0.020\,68\,(6)^f      &   \\[2pt]
$4F_{7/2}$  &   0.020\,170\,4\,(7)      &  0.020\,193\,(5)        &  0.020\,249\,(5)      &  0.020\,450\,(3)       &  0.021\,137\,(2)  \\
            &   0.020\,169\,90\,(2)^f   &  0.020\,192\,(3)^f      &  0.020\,25\,(2)^f     &  0.020\,4\,(2)^f       &   \\[2pt]
$5F_{7/2}$  &   0.020\,795\,(2)         &  0.020\,819\,(6)        &  0.020\,882\,(9)      &  0.021\,115\,(3)       &  0.021\,899\,5\,(4)  \\
            &   0.020\,794\,73\,(3)^f   &  0.020\,820\,(3)^f      &  0.020\,89\,(3)^f     &  0.021\,1\,(2)^f       &   \\[2pt]
$5G_{7/2}$  &  -0.012\,861\,(1)         & -0.012\,859\,(6)        & -0.012\,846\,(6)      & -0.012\,805\,(2)       & -0.012\,695\,(3)  \\
            &  -0.012\,859\,159\,(3)^f  & -0.012\,854\,6\,(3)^f   & -0.012\,843\,(3)^f    & -0.012\,81\,(2)^f      &   \\[2pt]
$5G_{9/2}$  &   0.012\,142\,(1)       &  0.012\,148\,(4)          &  0.012\,164\,(8)      &  0.012\,224\,4\,(9)    &  0.012\,441\,(3)  \\
            &   0.012\,140927\,(8)^f  &  0.012\,1475\,(9)^f       &  0.012\,165\,(8)^f    &  0.012\,23\,(6)^f      &   \\[2pt]
\end{tabular}
\end{ruledtabular}
\end{minipage}
\end{center}
$^a$ Jentschura and Mohr 2001 \cite{jentschura:01:pra},\\
$^b$ Jentschura 2004 \cite{jentschura:04:se},\\
$^c$ Indelicato and Mohr 1998 \cite{indelicato:98},\\
$^d$ Mohr and Kim 1992 \cite{mohr:92:a},\\
$^e$ Mohr 1992 \cite{mohr:92:b},\\
$^f$ $\Za$ expansion.
\end{table*}
\endgroup

\section{Supplemental material}

Tables~\ref{tab:s}-\ref{tab:g}.

\begin{table*}
\begin{center}
\caption{The one-loop electron self-energy correction for $nS$ states, in terms of $F(Z\alpha)$.
\label{tab:s}
}
\begin{ruledtabular}
\begin{tabular}{lw{3.12}w{3.12}w{3.12}w{3.12}w{3.12}}
\multicolumn{1}{l}{$Z$}
				                           &  \multicolumn{1}{c}{$1S_{1/2}$}
    				                                &  \multicolumn{1}{c}{$2S_{1/2}$}
    				                                &  \multicolumn{1}{c}{$3S_{1/2}$}
    				                                &  \multicolumn{1}{c}{$4S_{1/2}$}
    				                                &  \multicolumn{1}{c}{$5S_{1/2}$}
\\[2pt]
\hline\\[-7pt]
%
%
  1 &       10.31679350\,(67)       &       10.54682501\,(37)       &       10.60561402\,(23)       &       10.6293882\,(15)        &       10.6413494\,(19)         \\
  5 &        6.251627047\,(37)      &        6.48486021\,(35)       &        6.54338581\,(19)       &        6.56675872\,(66)       &        6.5783903\,(28)         \\
 10 &        4.654162327\,(31)      &        4.89441610\,(21)       &        4.95241028\,(38)       &        4.97491952\,(81)       &        4.9858390\,(13)         \\
 15 &        3.801410796\,(36)      &        4.050875801\,(65)      &        4.10822963\,(19)       &        4.12962767\,(22)       &        4.13962145\,(30)        \\
 20 &        3.246255619\,(18)      &        3.506647698\,(49)      &        3.56330226\,(21)       &        3.58340268\,(26)       &        3.59230900\,(40)        \\
 25 &        2.850104204\,(25)      &        3.122959418\,(59)      &        3.17887534\,(12)       &        3.19751576\,(17)       &        3.20519184\,(48)        \\
 30 &        2.552015178\,(18)      &        2.838838626\,(44)      &        2.89397986\,(15)       &        2.91099958\,(17)       &        2.91730503\,(30)        \\
 35 &        2.319976122\,(13)      &        2.622336143\,(37)      &        2.676659022\,(98)      &        2.69188472\,(21)       &        2.69667021\,(39)        \\
 40 &        2.135228445\,(15)      &        2.454829062\,(29)      &        2.50827295\,(14)       &        2.52150620\,(17)       &        2.52460453\,(65)        \\
 45 &        1.985943816\,(25)      &        2.324689560\,(33)      &        2.37716661\,(13)       &        2.38817321\,(21)       &        2.38939037\,(78)        \\
 50 &        1.864274384\,(44)      &        2.224337360\,(60)      &        2.27572253\,(13)       &        2.28421906\,(30)       &        2.28332545\,(46)        \\
 55 &        1.764830438\,(30)      &        2.148726571\,(47)      &        2.19884433\,(22)       &        2.20448374\,(30)       &        2.2012039\,(12)         \\
 60 &        1.683835894\,(26)      &        2.094517654\,(29)      &        2.14312592\,(93)       &        2.14547904\,(78)       &        2.1394783\,(13)         \\
 65 &        1.618636538\,(26)      &        2.059610843\,(24)      &        2.1063792\,(10)        &        2.10491073\,(58)       &        2.0957783\,(12)         \\
 70 &        1.567407614\,(28)      &        2.042891286\,(27)      &        2.0873720\,(10)        &        2.08140887\,(50)       &        2.06863668\,(93)        \\
 75 &        1.528984235\,(32)      &        2.044114703\,(33)      &        2.08570256\,(99)       &        2.07439057\,(53)       &        2.05734367\,(63)        \\
 80 &        1.502777576\,(41)      &        2.063906325\,(46)      &        2.10178156\,(96)       &        2.08402489\,(63)       &        2.06190080\,(79)        \\
 85 &        1.488762625\,(57)      &        2.103875899\,(70)      &        2.13692132\,(92)       &        2.11129572\,(75)       &        2.0830644\,(34)         \\
 90 &        1.487541918\,(94)      &        2.16688340\,(13)       &        2.19356045\,(89)       &        2.15818487\,(86)       &        2.1225078\,(64)         \\
 95 &        1.50051216\,(19)       &        2.25753959\,(25)       &        2.27569603\,(84)       &        2.22803536\,(94)       &        2.183130\,(17)          \\
100 &        1.53019940\,(41)       &        2.38312137\,(57)       &        2.38968015\,(78)       &        2.3262288\,(10)        &        2.2696851\,(88)         \\
\end{tabular}
\end{ruledtabular}
\end{center}
\end{table*}

\begin{table*}
\begin{center}
\caption{The one-loop electron self-energy correction for $nP_{1/2}$ states, in terms of $F(Z\alpha)$.
\label{tab:p1}
}
\begin{ruledtabular}
\begin{tabular}{lw{3.12}w{3.12}w{3.12}w{3.12}w{3.12}}
\multicolumn{1}{l}{$Z$}
				                           &  \multicolumn{1}{c}{$2P_{1/2}$}
    				                                &  \multicolumn{1}{c}{$3P_{1/2}$}
    				                                &  \multicolumn{1}{c}{$4P_{1/2}$}
    				                                &  \multicolumn{1}{c}{$5P_{1/2}$}
\\[2pt]
\hline\\[-7pt]
%
  1 &       -0.1263952\,(12)        &       -0.11545813\,(77)       &       -0.1104243\,(21)        &       -0.1076452\,(46)         \\
  5 &       -0.1227749\,(37)        &       -0.1112590\,(23)        &       -0.1060150\,(55)        &       -0.1031389\,(84)         \\
 10 &       -0.1148510\,(12)        &       -0.10204794\,(45)       &       -0.0963403\,(52)        &       -0.0932432\,(39)         \\
 15 &       -0.1045514\,(13)        &       -0.09005188\,(24)       &       -0.08373047\,(30)       &       -0.0803447\,(25)         \\
 20 &       -0.09251917\,(61)       &       -0.07601611\,(49)       &       -0.06897521\,(28)       &       -0.06525520\,(19)        \\
 25 &       -0.07906630\,(49)       &       -0.06030322\,(61)       &       -0.05246418\,(52)       &       -0.04838010\,(59)        \\
 30 &       -0.06433009\,(50)       &       -0.04308065\,(55)       &       -0.03438461\,(70)       &       -0.02992012\,(56)        \\
 35 &       -0.04834202\,(37)       &       -0.02439225\,(37)       &       -0.01479561\,(73)       &       -0.00994697\,(48)        \\
 40 &       -0.03104994\,(27)       &       -0.00418941\,(73)       &        0.00633664\,(62)       &        0.0115613\,(11)         \\
 45 &       -0.01233032\,(19)       &        0.01765678\,(58)       &        0.02912681\,(49)       &        0.0347044\,(13)         \\
 50 &        0.00801217\,(13)       &        0.04135515\,(36)       &        0.05376790\,(40)       &        0.05965934\,(31)        \\
 55 &        0.030252849\,(84)      &        0.06720145\,(41)       &        0.08053801\,(34)       &        0.0866872\,(12)         \\
 60 &        0.054763228\,(55)      &        0.09559624\,(43)       &        0.10981526\,(30)       &        0.1161415\,(16)         \\
 65 &        0.082037441\,(35)      &        0.12707157\,(42)       &        0.14210258\,(28)       &        0.1484953\,(18)         \\
 70 &        0.112732496\,(25)      &        0.16233245\,(33)       &        0.1780662\,(14)        &        0.1843733\,(15)         \\
 75 &        0.147729298\,(26)      &        0.20232055\,(36)       &        0.2185915\,(14)        &        0.2246079\,(14)         \\
 80 &        0.188226182\,(43)      &        0.24830834\,(31)       &        0.2648723\,(15)        &        0.2703167\,(21)         \\
 85 &        0.23588566\,(13)       &        0.30204782\,(29)       &        0.3185441\,(17)        &        0.3230272\,(39)         \\
 90 &        0.29307219\,(17)       &        0.36600996\,(34)       &        0.3818999\,(22)        &        0.3848743\,(73)         \\
 95 &        0.36325326\,(31)       &        0.44378215\,(51)       &        0.4582486\,(30)        &        0.458928\,(13)          \\
100 &        0.45171056\,(58)       &        0.54076699\,(85)       &        0.5525394\,(39)        &        0.549762\,(24)          \\
\end{tabular}
\end{ruledtabular}
\end{center}
\end{table*}

\begin{table*}
\begin{center}
\caption{The one-loop electron self-energy correction for $nP_{3/2}$ states, in terms of $F(Z\alpha)$.
\label{tab:p1}
}
\begin{ruledtabular}
\begin{tabular}{lw{3.12}w{3.12}w{3.12}w{3.12}w{3.12}}
\multicolumn{1}{l}{$Z$}
				                           &  \multicolumn{1}{c}{$2P_{3/2}$}
    				                                &  \multicolumn{1}{c}{$3P_{3/2}$}
    				                                &  \multicolumn{1}{c}{$4P_{3/2}$}
    				                                &  \multicolumn{1}{c}{$5P_{3/2}$}
\\[2pt]
\hline\\[-7pt]
%
  1 &        0.1234980\,(21)        &        0.1344137\,(17)        &        0.1394395\,(16)        &        0.1422153\,(45)         \\
  5 &        0.1256233\,(21)        &        0.1367939\,(15)        &        0.1419087\,(15)        &        0.1447242\,(31)         \\
 10 &        0.13035468\,(86)       &        0.14208588\,(26)       &        0.1473946\,(16)        &        0.1502973\,(13)         \\
 15 &        0.13656743\,(79)       &        0.14903556\,(19)       &        0.15459731\,(26)       &        0.15761275\,(65)        \\
 20 &        0.14383908\,(37)       &        0.15718575\,(40)       &        0.16304599\,(19)       &        0.16619080\,(19)        \\
 25 &        0.15192119\,(29)       &        0.16627224\,(50)       &        0.172470387\,(92)      &        0.17575927\,(26)        \\
 30 &        0.16064729\,(24)       &        0.17612783\,(55)       &        0.18270080\,(22)       &        0.18614624\,(19)        \\
 35 &        0.16990080\,(16)       &        0.18663871\,(33)       &        0.19362486\,(34)       &        0.19723880\,(10)        \\
 40 &        0.179594818\,(98)      &        0.19772817\,(58)       &        0.20516803\,(36)       &        0.20896224\,(44)        \\
 45 &        0.189663106\,(59)      &        0.20934353\,(42)       &        0.21728072\,(36)       &        0.22126804\,(45)        \\
 50 &        0.200053677\,(36)      &        0.22144963\,(23)       &        0.22993360\,(36)       &        0.23412901\,(51)        \\
 55 &        0.210724611\,(27)      &        0.23402526\,(33)       &        0.24311180\,(35)       &        0.24753174\,(21)        \\
 60 &        0.221641059\,(12)      &        0.24705904\,(44)       &        0.25681218\,(33)       &        0.26147594\,(23)        \\
 65 &        0.232772871\,(95)      &        0.26054766\,(61)       &        0.27104093\,(52)       &        0.27597111\,(29)        \\
 70 &        0.244092547\,(91)      &        0.27449373\,(79)       &        0.28581149\,(74)       &        0.29103461\,(82)        \\
 75 &        0.255573164\,(88)      &        0.2889031\,(10)        &        0.30114317\,(67)       &        0.3066905\,(15)         \\
 80 &        0.267186108\,(48)      &        0.3037831\,(12)        &        0.3170568\,(13)        &        0.3229649\,(30)         \\
 85 &        0.278898192\,(3)       &        0.3191386\,(13)        &        0.3335731\,(10)        &        0.3398839\,(80)         \\
 90 &        0.290667806\,(64)      &        0.3349663\,(12)        &        0.3507051\,(22)        &        0.3574633\,(62)         \\
 95 &        0.30243926\,(14)       &        0.35124712\,(87)       &        0.3684496\,(38)        &        0.375710\,(10)          \\
100 &        0.31413420\,(20)       &        0.36793220\,(16)       &        0.3867719\,(32)        &        0.394593\,(13)          \\
\end{tabular}
\end{ruledtabular}
\end{center}
\end{table*}

\begin{table*}
\begin{center}
\caption{The one-loop electron self-energy correction for $nD_{3/2}$ states, in terms of $F(Z\alpha)$.
\label{tab:d3}
}
\begin{ruledtabular}
\begin{tabular}{lw{3.12}w{3.12}w{3.12}w{3.12}w{3.12}}
\multicolumn{1}{l}{$Z$}
    				                                &  \multicolumn{1}{c}{$3D_{3/2}$}
    				                                &  \multicolumn{1}{c}{$4D_{3/2}$}
    				                                &  \multicolumn{1}{c}{$5D_{3/2}$}
\\[2pt]
\hline\\[-7pt]
%
  1 &       -0.0430187\,(22)        &       -0.0410066\,(29)        &       -0.0398608\,(38)         \\
  5 &       -0.04292701\,(98)       &       -0.0409033\,(31)        &       -0.0397525\,(32)         \\
 10 &       -0.0427075\,(27)        &       -0.0406609\,(35)        &       -0.0394963\,(44)         \\
 15 &       -0.0424068\,(38)        &       -0.0403196\,(22)        &       -0.0391298\,(53)         \\
 20 &       -0.0420204\,(24)        &       -0.0398802\,(12)        &       -0.03866266\,(69)        \\
 25 &       -0.0415540\,(11)        &       -0.03934426\,(65)       &       -0.03808744\,(24)        \\
 30 &       -0.0410040\,(25)        &       -0.03870380\,(45)       &       -0.03739636\,(15)        \\
 35 &       -0.0403623\,(15)        &       -0.03794651\,(45)       &       -0.03657495\,(40)        \\
 40 &       -0.03961784\,(65)       &       -0.03705648\,(50)       &       -0.03560437\,(45)        \\
 45 &       -0.03875914\,(78)       &       -0.03601460\,(53)       &       -0.03446263\,(46)        \\
 50 &       -0.03777035\,(80)       &       -0.03479869\,(52)       &       -0.03312413\,(47)        \\
 55 &       -0.03663449\,(57)       &       -0.03338331\,(46)       &       -0.03155963\,(47)        \\
 60 &       -0.03533042\,(73)       &       -0.03173954\,(38)       &       -0.02973576\,(46)        \\
 65 &       -0.03383567\,(70)       &       -0.02983449\,(30)       &       -0.0276140\,(14)         \\
 70 &       -0.03212545\,(73)       &       -0.0276303\,(20)        &       -0.0251525\,(15)         \\
 75 &       -0.03017023\,(80)       &       -0.0250855\,(20)        &       -0.0223011\,(19)         \\
 80 &       -0.02793704\,(89)       &       -0.0221507\,(21)        &       -0.0190038\,(31)         \\
 85 &       -0.02538843\,(94)       &       -0.0187702\,(25)        &       -0.0151960\,(60)         \\
 90 &       -0.02248185\,(90)       &       -0.0148803\,(35)        &       -0.010804\,(12)          \\
 95 &       -0.01916907\,(69)       &       -0.0104081\,(50)        &       -0.005744\,(23)          \\
100 &       -0.01539577\,(26)       &       -0.0052707\,(67)        &        0.000079\,(43)          \\
\end{tabular}
\end{ruledtabular}
\end{center}
\end{table*}

\begin{table*}
\begin{center}
\caption{The one-loop electron self-energy correction for $nD_{5/2}$ states, in terms of $F(Z\alpha)$.
\label{tab:d5}
}
\begin{ruledtabular}
\begin{tabular}{lw{3.12}w{3.12}w{3.12}w{3.12}w{3.12}}
\multicolumn{1}{l}{$Z$}
    				                                &  \multicolumn{1}{c}{$3D_{5/2}$}
    				                                &  \multicolumn{1}{c}{$4D_{5/2}$}
    				                                &  \multicolumn{1}{c}{$5D_{5/2}$}
\\[2pt]
\hline\\[-7pt]
%
  1 &        0.0403166\,(10)        &        0.0423291\,(10)        &        0.0434761\,(22)         \\
  5 &        0.04043227\,(60)       &        0.0424598\,(40)        &        0.0436123\,(59)         \\
 10 &        0.0407339\,(21)        &        0.0428009\,(17)        &        0.0439733\,(18)         \\
 15 &        0.0411800\,(31)        &        0.0433083\,(15)        &        0.0445063\,(31)         \\
 20 &        0.0417506\,(20)        &        0.04395918\,(51)       &        0.0451955\,(15)         \\
 25 &        0.04243887\,(93)       &        0.04474925\,(10)       &        0.04603137\,(22)        \\
 30 &        0.0432408\,(19)        &        0.04567522\,(12)       &        0.047013577\,(76)       \\
 35 &        0.0441536\,(11)        &        0.04673599\,(32)       &        0.04814108\,(12)        \\
 40 &        0.04517537\,(54)       &        0.04793174\,(50)       &        0.04941502\,(21)        \\
 45 &        0.04630663\,(61)       &        0.04926333\,(59)       &        0.05083707\,(29)        \\
 50 &        0.04754638\,(60)       &        0.05073184\,(59)       &        0.05240924\,(36)        \\
 55 &        0.04889435\,(44)       &        0.05233814\,(52)       &        0.05413269\,(37)        \\
 60 &        0.05034847\,(47)       &        0.05408257\,(40)       &        0.05600857\,(35)        \\
 65 &        0.05190829\,(40)       &        0.05596468\,(26)       &        0.05803675\,(51)        \\
 70 &        0.05357112\,(33)       &        0.05798300\,(65)       &        0.0602157\,(18)         \\
 75 &        0.05533370\,(27)       &        0.0601343\,(14)        &        0.0625440\,(13)         \\
 80 &        0.05719170\,(22)       &        0.0624150\,(12)        &        0.0650167\,(19)         \\
 85 &        0.05913955\,(18)       &        0.0648182\,(28)        &        0.0676274\,(58)         \\
 90 &        0.06117035\,(16)       &        0.0673369\,(17)        &        0.0703652\,(47)         \\
 95 &        0.06327573\,(14)       &        0.0699603\,(17)        &        0.0732217\,(93)         \\
100 &        0.06544582\,(11)       &        0.0726767\,(15)        &        0.076182\,(14)          \\
\end{tabular}
\end{ruledtabular}
\end{center}
\end{table*}

\begin{table*}
\begin{center}
\caption{The one-loop electron self-energy correction for $nF$ states, in terms of $F(Z\alpha)$.
\label{tab:f}
}
\begin{ruledtabular}
\begin{tabular}{lw{3.12}w{3.12}w{3.12}w{3.12}w{3.12}}
\multicolumn{1}{l}{$Z$}
    				                                &  \multicolumn{1}{c}{$4F_{5/2}$}
    				                                &  \multicolumn{1}{c}{$5F_{5/2}$}
    				                                &  \multicolumn{1}{c}{$4F_{7/2}$}
    				                                &  \multicolumn{1}{c}{$5F_{7/2}$}
\\[2pt]
\hline\\[-7pt]
%
  1 &       -0.0214977\,(14)        &       -0.0208740\,(19)        &        0.02017035\,(65)       &        0.0207953\,(15)         \\
  5 &       -0.0214796\,(42)        &       -0.0208571\,(53)        &        0.0201926\,(47)        &        0.0208192\,(60)         \\
 10 &       -0.0214379\,(44)        &       -0.0208094\,(98)        &        0.0202487\,(53)        &        0.0208819\,(90)         \\
 15 &       -0.0213817\,(25)        &       -0.02074480\,(85)       &        0.0203372\,(20)        &        0.0209829\,(17)         \\
 20 &       -0.0213094\,(26)        &       -0.0206675\,(26)        &        0.0204502\,(30)        &        0.0211150\,(26)         \\
 25 &       -0.0212327\,(44)        &       -0.0205805\,(29)        &        0.0205859\,(29)        &        0.0212723\,(23)         \\
 30 &       -0.0211397\,(38)        &       -0.0204767\,(13)        &        0.0207508\,(34)        &        0.02145555\,(88)        \\
 35 &       -0.0210355\,(30)        &       -0.02036000\,(92)       &        0.0209331\,(27)        &        0.02166476\,(53)        \\
 40 &       -0.0209204\,(21)        &       -0.02023031\,(65)       &        0.0211366\,(19)        &        0.02189948\,(36)        \\
 45 &       -0.0207941\,(13)        &       -0.02008692\,(50)       &        0.0213611\,(12)        &        0.02215963\,(32)        \\
 50 &       -0.02065620\,(67)       &       -0.01992881\,(43)       &        0.0216066\,(29)        &        0.02244530\,(33)        \\
 55 &       -0.0205064\,(27)        &       -0.01975463\,(39)       &        0.0218729\,(22)        &        0.02275674\,(37)        \\
 60 &       -0.0203429\,(20)        &       -0.01956276\,(38)       &        0.0221602\,(16)        &        0.02309435\,(39)        \\
 65 &       -0.0201648\,(14)        &       -0.0193507\,(29)        &        0.02246792\,(52)       &        0.02345848\,(74)        \\
 70 &       -0.0199705\,(50)        &       -0.0191176\,(29)        &        0.02279765\,(67)       &        0.0238495\,(19)         \\
 75 &       -0.0197596\,(44)        &       -0.0188603\,(30)        &        0.0231497\,(36)        &        0.0242688\,(15)         \\
 80 &       -0.0195296\,(40)        &       -0.0185766\,(39)        &        0.0235233\,(31)        &        0.0247160\,(32)         \\
 85 &       -0.0192787\,(37)        &       -0.0182639\,(60)        &        0.0239191\,(25)        &        0.0251931\,(27)         \\
 90 &       -0.0190051\,(38)        &       -0.017920\,(10)         &        0.0243376\,(21)        &        0.0257001\,(62)         \\
 95 &       -0.0187067\,(43)        &       -0.017541\,(19)         &        0.0247791\,(19)        &        0.0262349\,(47)         \\
100 &       -0.0183813\,(50)        &       -0.017126\,(32)         &        0.0252442\,(25)        &        0.0268017\,(77)         \\
\end{tabular}
\end{ruledtabular}
\end{center}
\end{table*}

\begin{table*}
\begin{center}
\caption{The one-loop electron self-energy correction for $nG$ states, in terms of $F(Z\alpha)$.
\label{tab:g}
}
\begin{ruledtabular}
\begin{tabular}{lw{3.12}w{3.12}w{3.12}w{3.12}w{3.12}}
\multicolumn{1}{l}{$Z$}
    				                                &  \multicolumn{1}{c}{$5G_{7/2}$}
    				                                &  \multicolumn{1}{c}{$5G_{9/2}$}
\\[2pt]
\hline\\[-7pt]
%
  1 &       -0.0128605\,(13)        &        0.0121415\,(11)         \\
  5 &       -0.0128586\,(60)        &        0.0121484\,(42)         \\
 10 &       -0.0128458\,(59)        &        0.0121641\,(76)         \\
 15 &       -0.0128240\,(22)        &        0.0121902\,(55)         \\
 20 &       -0.0128050\,(20)        &        0.01222435\,(88)        \\
 25 &       -0.0127802\,(25)        &        0.0122683\,(24)         \\
 30 &       -0.0127526\,(22)        &        0.0123182\,(29)         \\
 35 &       -0.0127273\,(35)        &        0.0123733\,(23)         \\
 40 &       -0.0126954\,(32)        &        0.0124405\,(30)         \\
 45 &       -0.0126608\,(28)        &        0.0125107\,(26)         \\
 50 &       -0.0126236\,(23)        &        0.0125873\,(21)         \\
 55 &       -0.0125839\,(17)        &        0.0126702\,(16)         \\
 60 &       -0.0125416\,(13)        &        0.0127593\,(11)         \\
 65 &       -0.01249669\,(83)       &        0.01285467\,(70)        \\
 70 &       -0.0124479\,(59)        &        0.0129566\,(23)         \\
 75 &       -0.0123980\,(58)        &        0.0130637\,(47)         \\
 80 &       -0.0123453\,(59)        &        0.0131782\,(42)         \\
 85 &       -0.0122898\,(66)        &        0.0132992\,(36)         \\
 90 &       -0.0122315\,(84)        &        0.0134252\,(76)         \\
 95 &       -0.012170\,(12)         &        0.0135595\,(53)         \\
100 &       -0.012106\,(17)         &        0.0137005\,(52)         \\
\end{tabular}
\end{ruledtabular}
\end{center}
\end{table*}

\end{document}